\documentclass{emulateapj}
\usepackage{amsmath}
\usepackage{lscape}
\usepackage{rotating}
\begin{document}
\title{Serendipitous Discovery of an Overdensity of Ly$\alpha$ Emitters at $z\sim$ 4.8 in the Cl1604 Supercluster Field}
\author{Lemaux, B.C., Lubin, L.M., Sawicki, M.\altaffilmark{1,2}, Martin, C.\altaffilmark{2,5}, Lagattuta, D.J., Gal, R.R.\altaffilmark{3}, Kocevski, D., Fassnacht, C.D. \& Squires, G. K.\altaffilmark{4}}
\affil{Department of Physics, University of California, Davis, 1 Shields Avenue, Davis, CA 95616, USA}
\email{lemaux@physics.ucdavis.edu}
\altaffiltext{1}{Department of Astronomy and Physics, Saint Mary's University, 923 Robie Street, Halifax, Nova Scotia, B3H 3C3, Canada}
\altaffiltext{2}{Department of Physics, University of California, Santa Barbara, Santa Barbara, CA 93106, USA}
\altaffiltext{3}{University of Hawai'i, Institute for Astronomy, 2680 Woodlawn Drive, Honolulu, HI 96822, USA}
\altaffiltext{4}{California Institute of Technology, M/S 220-6, 1200 E. California Blvd., Pasadena, CA 91125, USA}
\altaffiltext{5}{Packard Fellow}
\begin{abstract}
We present results of a spectroscopic search for Ly$\alpha$ emitters (LAEs) in the Cl1604 supercluster field using
the extensive spectroscopic Keck/DEep Imaging Multi-Object Spectrograph database taken as part of the 
Observations of Redshift Evolution in Large
Scale Environments (ORELSE) survey. A total of 12 slitmasks were observed and inspected in the Cl1604 field, spanning a survey
volume of $1.365 \times 10^{4}$ co-moving $\rm{Mpc}^{3}$. We find a total of 17 high redshift ($4.39\leq z\leq5.67$) 
LAE candidates down to a limiting flux of $1.9\times10^{-18}$ ergs s$^{-1}$ cm$^{-2}$  ($L($Ly$\alpha)=4.6\times10^{41}$ ergs $\rm{s}^{-1}$ 
or $\sim$0.1$L_{\ast}$ at $z\sim5$), 13 of which 
we classify as high quality. The resulting LAE number density is nearly double that of LAEs found in the Subaru deep field at $z\sim4.9$ and 
nearly an order of magnitude higher than in other surveys of LAEs at similar redshifts, an excess that is essentially independent of 
LAE luminosity. We also report on the discovery of two possible LAE group structures at $z\sim4.4$ and $z\sim4.8$ and investigate the effects
of cosmic variance of LAEs on our results.  Fitting a simple truncated single Gaussian 
model to a composite spectrum of the 13 high quality LAE candidates, we find a best-fit stellar velocity dispersion of 136 km $\rm{s}^{-1}$. Additionally, we see modest evidence of 
a second peak in the composite spectrum, possibly caused by galactic outflows, offset from the main velocity centroid of the LAE population by $\sim$440 km $\rm{s}^{-1}$ as well as evidence 
for a non-trivial Ly$\alpha$ escape fraction. We find an average star formation rate density (SFRD) of $\sim5\times10^{-3}$ $\rm{M}_{\odot}$ $\rm{yr}^{-1}$ $\rm{Mpc}^{-3}$ with moderate evidence for
negative evolution in the SFRD from $z\sim4.6$ to $z\sim5.7$. By simulating the statistical flux-loss due to our observational setup we measure a best-fit luminosity function characterized by $\Phi_{\ast}L_{\ast}=2.2^{+3.9}_{-1.3}\times10^{39}$ ergs s$^{-1}$ Mpc$^{-3}$ for $\alpha$=-1.6, generally consistent with measurements from other surveys at similar epochs.
Finally, we investigate any possible effects from weak or strong gravitational lensing induced by the foreground supercluster, 
finding that our LAE candidates are minimally affected by lensing processes. 

\end{abstract}
\section{Introduction}

While Ly$\alpha$ emitters (LAEs) have been sought for nearly 40 years, designing and 
implementing surveys capable of detecting large unbiased populations of these objects have 
proven difficult. Due to the extreme faintness of the population and technological limitations, 
the searches pioneered by Davis et al.\ in the 1970s (Davis \& Wilkinson 1974; Partridge 1974) established what 
would later be a theme for such surveys: constraints on galaxy populations and cosmological 
parameters through a dearth of detections. At that time little was known about the 
properties of high-redshift galaxies, with the observational distinction between LAEs and a second high-redshift
star-forming population, Lyman break galaxies (LBGs), not yet possible. This ignorance about the fundamental 
differences in the properties of the two types of high-redshift galaxies resulted in the 
grouping of both galaxy populations into a single category: Primeval Galaxies (PGs). While 
early theoretical modeling (see Davis 1980) predicted the density of PGs to be $\gtrsim$10000 per 
$\rm{deg}^{2}$ at high redshift (z $>$ 3), early searches for PGs (Koo \& Kron 1980; Saulson \& Boughn et al. 1982;
Boughn et al. 1986; Pritchet \& Hartwick 1987, 1990; Elston et al. 1989; de Propris et al. 1993; Thompson et al. 1995; Thompson \& Djorgovski 1995) 
were unable to find any such objects. It was not until the mid-1990's with the 
searches of Steidel and collaborators that large populations of PGs were detected, almost  
exclusively of the LBG flavor (Steidel et al. 1996a, 1996b). 

The detection of LAEs has proven significantly more problematic than LBGs due to 
the difficulty of efficiently identifying the Ly$\alpha$ line in candidate galaxies. In addition,
the Ly$\alpha$ line is only observed in $\sim$25\% of high redshift star-forming galaxies
(Steidel et al. 2000; Shapley et al. 2003). Due to these difficulties, it is only in the 
past half-decade that techniques have been successfully developed and implemented 
to detect reasonably large numbers of LAEs. 

The most common technique in contemporary LAE searches is the use of
custom-made narrowband filters with bandpasses of 100 \AA\ or 
less, designed to collect light in windows of low atmospheric transmission. 
Imaging campaigns using such filters have been successfully undertaken in 
blank fields complemented by deep broadband photometry (Hu et al. 2004, hereafter H04; Ouchi et al. 2003, 2008, 
hereafter O03, O08; Rhoads et al. 2000; Malhotra \& Rhoads 2002) or in areas of suspected overdensities 
(Kurk et al. 2004; Miley et al. 2004; Venemans et al. 2004; Zheng et al. 2006; Overzier et al. 2008). 
While this technique has proven capable of detecting 
large numbers of LAEs, the populations detected may be inherently biased, 
due either to the small redshift windows probed, a bias intensified
by the high level of observed spatial clustering of LAEs,
or due to the large line equivalent widths (EWs) necessary to detect such objects.

An alternative is dedicated spectroscopic campaigns in blank fields (Crampton \& 
Lilly 1999; Martin \& Sawicki 2004, hereafter MS04; Tran et al. 2004, hereafter T04; Martin et al. 2008, hereafter M08), yielding samples of 
LAEs complementary to photometric searches. 
While narrowband imaging surveys provide large samples as a result of 
their ability to probe large volumes in relatively short periods of time, 
the increased sky noise due to the large filter bandpass ($\sim$100\AA) 
relative to a ``typical" Ly$\alpha$ emission width (10-20 \AA\ full-width at half-maximum, FWHM) makes it 
difficult to probe deep into the LAE luminosity function. As a result the line luminosities of 
galaxies detected in these surveys are usually at or above $L_{\ast}$. By 
dispersing the night sky background so that the emission line has only to 
exceed the background over the natural width of the 
line rather than over $\sim$ 100 \AA, spectroscopic surveys for LAEs become much more efficient 
probes of sub-$L_{\ast}$ galaxies at high redshift.

The difficulty with such observations is that spectroscopy probes a significantly smaller 
area on the sky than narrowband techniques, with the area reduced by the ratio of the slit area 
to the telescope field of view (see discussion in M08). The early dedicated searches 
of T04 and MS04 suffered from this limitation, covering 17.6 $\rm{arcmin}^{2}$ and 5.1 $\rm{arcmin}^{2}$ respectively. 
Along with the small spectral bandpasses designed to fit in atmospheric transmission windows, this effect severely 
limited the volume probed by such surveys and 
as a result no LAEs were detected. It was not until the recent search of M08, using similar techniques but with a significant increase in sensitivity and field of view, that LAEs were discovered exclusively through dedicated spectroscopic techniques. These 
results demonstrate the necessity of large volume searches to effectively detect and analyze 
populations of LAEs. 

With the recent use of multi-object spectrographs for large surveys of galaxies at intermediate redshift (e.g., DEEP2, VVDS) 
it has become possible to obtain deep, high resolution spectra of large patches of blank sky and move beyond 
single serendipitous discoveries of LAEs (Franx et al. 1997; Dawson et al. 2002; Stern et al. 2005) to 
statistical samples of high-redshift emission line galaxies (Sawicki et al. 2008; hereafter S08). With this in mind, we have 
searched the extensive (3.214 $\rm{arcmin}^{2}$, 1.365$\times10^{4}$ $\rm{Mpc}^{3}$) spectroscopic database 
of the Cl1604 supercluster at $z$ $\sim$ 0.9 (Gal et al. 2008, hereafter G08). This structure is studied as part of the 
Observations of Redshift Evolution in Large 
Scale Environments (ORELSE) survey (Lubin et al. 2009), an ongoing 
multi-wavelength campaign mapping out the environmental effects on galaxy 
evolution in the large scale structures surrounding 20 known clusters at moderate redshift ($0.6 \leq z \leq 1.3$). 
While the angular coverage is moderate compared to other such surveys of LAEs, the Cl1604 data 
have the advantage of large spectral coverage (see Section 2) and deep observations on the Keck 10-m telescope, which allow us 
to probe down to unprecedented levels in the luminosity function ($\sim$0.1$L_{\ast}$ at $z\sim5$). As a result we find 17 LAE candidates in 
our moderately sized volume, almost all of which are fainter than the characteristic luminosity at $z\sim5$. These detections allow us to 
place some of the first constraints on the properties of low luminosity galaxies at high 
redshift, including implications for this population's role in the reionization of the universe.  

The remainder of the paper is organized as follows: Section 2 describes the spectral data and our 
selection process. Section 3 describes tests to validate
our high redshift LAE candidates. Section 4 includes a discussion of other properties, 
such as photometric limits, line equivalent 
widths, velocity profiles, and star formation rates (SFRs) of the LAE candidates. In Section 5 we describe 
the number density and luminosity function of our LAE candidates as well as the effects of LAE clustering and cosmic
variance. In addition, since these data were taken in an area of 
the sky with a rare, massive structure in the foreground, we also discuss in Section 5 any possible contributions from 
gravitational lensing. Section 6 summarizes our results. Throughout this paper we use the 
concordance $\Lambda$CDM cosmology with 
$H_{0}$ = 70 km s$^{-1}$, $\Omega_{\Lambda}$ = 0.7, and $\Omega_{M}$ = 0.3. At $z=4.8$, the median redshift of our sample, 
the age of the universe is 1.2 Gyr and the angular scale is 6.41 kpc arcsec$^{-1}$, with 621 Myr elapsing between 
$z=6.4$ and $z=4.1$, the redshift range of LAEs to which our spectral coverage is sensitive. All EW measurements
are given in the rest frame and all magnitudes are given in the AB system (Oke \& Gunn 1983; Fukugita et al. 1996).

\section{Data}

The first target of the ORELSE survey, and the subject of study 
in this paper, is the Cl1604 field, containing the Cl1604 supercluster at z = 0.9: a massive collection 
of eight or more constituent groups and clusters spanning 13 $h^{-1}$ 
comoving Mpc in the transverse dimensions and nearly 100 $h^{-1}$ comoving Mpc 
in the radial dimension (see G08 for the coordinates and velocity centroids of the 
clusters that comprise the Cl1604 supercluster). The data on this structure include Very Large Array (B-array, 20 cm), 
\emph{Spitzer} IRAC (3.6/4.5/5.8/8.0 $\mu$m) and MIPS 24 $\mu$m imaging, archival Subaru V-band 
imaging, deep Palomar $r\arcmin$ $i\arcmin$ $z\arcmin$ $K_{s}$ imaging, 
a 17 pointing \emph{Hubble Space Telescope} ACS mosaic in F606W and F814W, and two deep (50 ks) \emph{Chandra} pointings.

In addition to the photometric data, an extensive spectroscopic campaign has been 
completed in the Cl1604 field to determine the rest-frame 
optical/UV spectral properties and redshifts of a large fraction of the 
constituent cluster members. Photometric data alone are not ideal for this 
purpose, as typical photometric redshift errors can span the line-of-sight 
extent of large scale structures such us Cl1604, leading to severe uncertainties 
in environmental indicators such as local density. To accurately 
quantify environmental effects, large spectroscopic coverage is essential in 
minimizing the effects of 
projections (see G08 for a more detailed discussion). 

To this end, 12 masks covering a large portion of the Cl1604 structure were 
observed with the DEep Imaging Multi-Object Spectrograph (DEIMOS; Faber et 
al. 2003) on the Keck II 10-m telescope between May 2003 and June 2007. 
The observations were taken with 1$\arcsec$ slits with the 1200 l mm$^{-1}$ grating, blazed 
at 7500 \AA, resulting in a pixel scale of 0.33 \AA\ pix$^{-1}$, a resolution of 
$\sim$ 1.7 \AA\  (68 km s$^{-1}$), and typical wavelength coverage of 6385 
\AA\ to 9015 \AA. Each DEIMOS mask contained between 80 and 130 individual 
slits with an average length of 9.9$\arcsec$, with 95\% having slit lengths between 
4.92$\arcsec$ and 14.88$\arcsec$. The slits in each mask combined for a total sky coverage 
of 0.2678 arcmin$^{2}$ per mask, independent of the number of slits. The 
spectroscopic targets 
for these slits were selected based on the likelihood of being a cluster member, 
 determined through a series of color and magnitude selections (see G08). 
The masks were observed with differing total integration 
times, which varied depending on weather and seeing conditions, in order to achieve similar 
levels of redshift completeness of targeted galaxies. A differing 
number of 1800s exposures were stacked for each mask, with total integration 
times of 7200s to 14400s. 

The exposures for each mask were combined using the DEEP2 version of the 
\emph{spec2d} package (Davis et al. 2003)\footnote{See also http://astro.berkeley.edu/$\sim$cooper/deep/spec2d/}. 
This package combines the individual 
exposures of the slit mosaic and performs wavelength calibration, cosmic ray 
removal and sky subtraction on slit by slit basis, generating a processed two-dimensional spectrum 
for each slit. The \emph{spec2d} pipeline also generates a processed 
one-dimensional spectrum for each slit. This extraction creates a one-dimensional spectrum of the 
target, containing the summed flux at each wavelength in an optimized window. In all, 903 
total high quality (Q $\geq$ 3, see G08 for an explanation on the quality codes)
spectra were obtained, with 329 falling within 0.84 $\leq$ z $\leq$ 0.96, the adopted redshift 
range of the supercluster.

\subsection{Searching for Serendipitous Detections}

During the reduction process \emph{spec2d} also determines if any other peaks exist in the 
spatial profile of the slit that are distinct from the target. If such peaks 
exist, \emph{spec2d} does similar extractions at these spatial locations creating 
one-dimensional spectra for these non-targeted serendipitous detections (hereafter serendips). 
All serendipitous spectra generated in this manner were systematically 
inspected by one of us (RG) to determine whether these extractions contained genuine stellar or 
galactic signatures rather than instrumental or reduction artifacts. 

In addition to the \emph{spec2d} extraction algorithm for serendips, each mask was visually inspected
by two of the authors (BL and RG) independently to search for additional serendips using \emph{zspec}, a publicly available 
redshift measurement program developed by D. Magwick, M. Cooper, and N. Konidaris for the DEEP2 survey. In the few 
cases where an object was found by only
one of the authors or an object was assigned two separate redshifts, the slit was ``blindly" re-analyzed by a third author (DK)
and a consensus was reached on the validity and redshift of the serendip by all three of the authors. Once a serendip 
was found by eye and confirmed genuine, and if \emph{spec2d} had not detected it on the slit, a manual extraction was performed. This 
process involved re-running the \emph{spec2d} extraction routine on the two-dimensional spectrum with a 
centroid and FWHM determined by the spatial
location and extent of the serendips as measured in the two-dimensional spectrum. 
This new extraction was then inspected and analyzed using in \emph{zspec} to determine if the extraction window was properly centered and 
the aperture was properly matched to the spatial extent of the source. In the cases where a non-targeted object 
was detected by eye and \emph{spec2d} had correctly extracted the spectrum, the 
one-dimensional spectrum was displayed with 
\emph{zspec} and, if needed, any modifications to the centroid and FWHM were done iteratively. The 
redshift was determined by guessing the wavelength range of a feature (typically 
3727 \AA\  [OII], 3968 \AA\  CaH, 3934 \AA\  CaK, 4861 \AA\  H$\beta$, 5007 \AA\  
[OIII], or 6563 \AA\  H$\alpha$), which allows \emph{zspec} to 
determine the best-fit redshift through an iterative $\chi^{2}$ minimization algorithm. All serendips found 
in the Cl1604 spectral data were found through visual inspection, only 30\% of which were also 
detected and extracted by \emph{spec2d}. The small fraction of serendips detected by \emph{spec2d} is not surprising
as most galaxies discovered serendipitously were faint emission-line objects and \emph{spec2d} requires either 
a continuum or several bright emission features to recognize and extract the spectrum of a second object on the slit.

Of the 167 serendips found in this manner, 122 were associated with the 
previously mentioned lower redshift ($z$ $<$ 1 for our spectral setup) nebular 
emission or stellar absorption lines. The remaining 45 objects were 
associated with either (a) low signal-to-noise ratio (S/N) features making a redshift determination 
uncertain, (b) definite features obscured by poor sky reduction or other 
instrumental issues, or (c) a single feature, which in the absence of any other 
spectral indicators makes redshift determination difficult, but not impossible 
(Kirby et al. 2007). It is the 39 galaxies which comprise category (c) that are
of interest for this paper. 

\subsection{Survey Volume}

The 12 DEIMOS masks observed in the field of the Cl1604 supercluster subtend a total angular 
area of 3.214 $\rm{arcmin}^{2}$, significantly smaller than the 200 $\rm{arcmin}^{2}$ covered by 
the dedicated IMACS Magellan LAE survey of M08 and smaller than even the 5.1 $\rm{arcmin}^{2}$ covered 
by MS04.  However, these surveys have limited volume due to their relatively small coverage in the 
line of sight dimension, with spectral ranges comparable to that of narrowband imaging surveys 
($\sim$ 100 \AA). The large spectral coverage (6400 \AA\ to 9000 
\AA) of the Cl1604 DEIMOS data allows for a competitive survey volume. The 12 
masks sample a volume of $1.70\times10^{4}$ co-moving $\rm{Mpc}^{3}$ between $z = 4.26$ and $z = 6.40$, slightly smaller than 
other contemporary blind spectroscopic searches for LAEs ($4.5\times10^{4}$ comoving $\rm{Mpc}^{3}$, M08; $6.9\times10^{4}$ comoving $\rm{Mpc}^{3}$, S08). 
However, this volume still does not approach the volume covered in narrowband imaging searches for LAEs such as LALA ($7.4\times 10^{5} \, 
\rm{Mpc}^{3}$; Rhoads et al. 2000; Rhoads \& Malhotra 2001), the Subaru Deep Field Search ($\sim 1\times10^{6} \, \rm{Mpc}^{3}$; O03), the 
Subaru \emph{XMM - Newton} Deep Survey (hereafter SXDF) ($\sim 1\times10^{6} \, \rm{Mpc}^{3}$; O08), 
or the search for LAE galaxies in the COSMOS field ($\sim 1.7\times10^{6} 
\, \rm{Mpc}^{3}$; Murayama et al. 2007). Though surveying a volume significantly smaller than that of 
narrowband imaging searches, the Cl1604 data has the advantage of 
probing much deeper in the luminosity function than such surveys, with a limiting luminosity of $L_{lim}=4.6\times10^{41}$ 
ergs $\rm{s}^{-1}$ at $z\sim5$, an order of magnitude dimmer than those of narrowband imaging surveys ($L_{lim}=4\times10^{42}$ 
ergs $\rm{s}^{-1}$ , Rhoads \& Malhotra 2001; $L_{lim}=3\times10^{42}$ ergs $\rm{s}^{-1}$ , O08; 
$L_{lim}=6.3\times10^{42}$ ergs $\rm{s}^{-1}$ , Murayama et al. 2007). The limiting Ly-$\alpha$ 
luminosity varies slightly (5\%-10\%) from mask to mask due to different integration times and seeing conditions; however, the limiting luminosity 
of $L_{lim}=4.6\times10^{41}$ ergs $\rm{s}^{-1}$ represents the \emph{brightest} limiting luminosity at $z\sim5$ of all 12 masks, meaning 
that an LAE at $z\la5$ with this luminosity would be detected in all masks as long as it fell relatively close to the center of a DEIMOS slit.

Three LAEs were detected by M08, representing the first successful detection of LAEs by a dedicated 
spectroscopic survey. Given the survey volume of M08 and the range of luminosities found in their 
survey, it is reasonable to assume that to detect at least one LAE with L$\geq L_{\ast}$ a survey 
volume of $1.5 \times 10^{4}$ $\rm{Mpc}^{3}$ is needed. This is consistent with the non-detections of 
T04 and MS04, which covered $6.13\times10^{3}$ $\rm{Mpc}^{3}$ and $1.1\times10^{3}$ $\rm{Mpc}^{3}$, 
respectively, and were sensitive to this depth. This limit, which excludes the effects of sample variance or any 
evolution in the LAE luminosity function between $z$ = 4.26 and $z$ = 6.4, 
places our survey right at the volume threshold necessary to detect a single LAE with L $>$ $L_{\ast}$.

\begin{figure}
\plotone{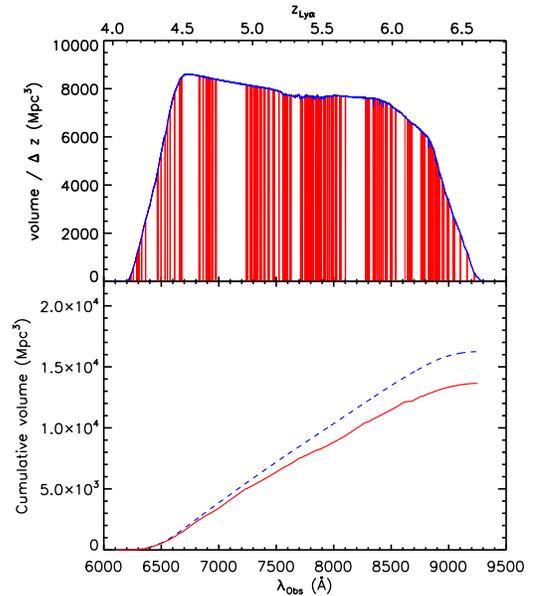}
\caption{The top panel shows differential volume (per unit redshift) as a function of wavelength for our 12 DEIMOS masks in the Cl1604 field. The vertical lines represent regions straddling 
bright night sky emission features. The lower panel shows the cumulative volume of the survey as a function of wavelength as corrected for the volume lost by the night sky emission lines (red solid line) and the uncorrected volume (blue dashed line).}
\label{fig:volume}
\end{figure}

To calculate the volume of the survey from the entire observable redshift range of the DEIMOS 
masks is, however, an overestimate; 
sky emission features render spectral regions of the data essentially unusable, necessitating bright line fluxes in order to 
exceed the sky noise. It is also tempting at this point to make a correction for the angular area of the slit lost by placing 
a relatively large lower redshift object (the targeted galaxy) in the center of each slit. However, as discussed in Section 3.2 
this portion of the slit is not rendered unusable by the target galaxy, as we find many serendips
and nearly one-third of our LAE candidate population at positions coincident with the spatial location of the targets. While it is extremely 
likely that the physics governing the observed luminosities at these locations differ from serendips discovered at other positions along the slit 
(the two most likely physical mechanisms are discussed briefly in Section 3.2), this portion of the slit can 
still be used to serendipitously detect galaxies and we therefore include it in the calculation of the volume 
probed by the survey. An estimate of the loss due to airglow lines is necessary, however, and must be
done on a slit-to-slit basis as the wavelength coverage of each slit is not uniform, but
depends on the position of the slit along the direction parallel to the dispersion on the
slitmask, and is further compounded by the non-uniformity of the spatial lengths of the slits. 
In order to properly account for the fractional volume lost by bright sky emission lines, we adopt 
an approach similar to the one taken in S08. For each two-dimensional slit file, the wavelength value of each pixel was determined from the 
\emph{spec2d} wavelength solution. Every pixel that was 
within $\pm$ 2$\sigma$ (calculated from the FWHM 1200 l mm$^{-1}$ resolution) of any bright 
night sky emission line was considered unusable. The high resolution of the 1200 l mm$^{-1}$
DEIMOS data allows for minimal losses in usable volume, losing only $\pm$ 1.7 \AA\ around each airglow line. 
Figure \ref{fig:volume} shows the usable elements 
of the data in the spectral dimension as well as the cumulative volume covered by the survey as a function of 
increasing wavelength. The volume calculated in this manner was $1.365 \times 10^{4}$ 
co-moving $\rm{Mpc}^{3}$, $\sim$ 20\% smaller than that determined by the more naive calculation. 

\subsection{Flux Calibration}

\begin{figure}
\plotone{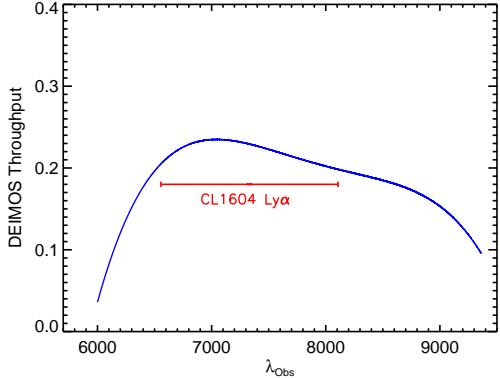}
\caption{A fifth-order Legendre polynomial fit to measured values of the throughput of DEIMOS for our spectral setup as a function of 
wavelength. This throughput includes loss from other optical elements and atmospheric transmission. For comparison the wavelength range of 
Cl1604 LAE candidates found in the data is plotted below the throughput curve. The wavelength range of LAE candidates encompasses the area of 
highest instrumental throughput.}
\label{fig:through}
\end{figure}

The DEIMOS spectra were flux calibrated using a fifth-order Legendre polynomial fit to time-averaged DEIMOS 1200 
l mm$^{-1}$ observations of spectrophotometric standard stars\footnote{See http://www.ucolick.org/$\sim$ripisc/results.html} 
taken between June 2002 and September 2002 (see Figure \ref{fig:through}). While the response
is known to vary as a function of time\footnote{See http://www.ucolick.org/$\sim$kai/DEEP/DEIMOS/summary.html}, 
it is a relatively small effect under photometric conditions ($\sim$5\%-10\%). 
As most of our data were taken under photometric conditions, we can safely ignore this variation. The throughput 
correction for each pixel is: 
\begin{equation}
f_{\lambda,i} [\frac{ergs}{cm^{2} \, s \, \AA}] = \frac{C_{i} \, D_{i} \, h \, c}{\pi 449^{2} \, \delta_{\lambda,i} \, t_{exp} \, \lambda_{c,i},}   
\label{eqn:flam}
\end{equation}

\noindent where $C_{i}$ are the raw counts in the $i$th pixel, $D_{i}$ is the throughput correction 
at the central wavelength of the $i$th pixel, 449 is half the effective Keck II mirror aperture in 
centimeters, $\delta_{\lambda,i}$ is the plate scale in the $i$th pixel in \AA\ pixel$^{-1}$, $t_{exp}$ 
is the effective exposure time\footnote{The effective exposure time is 3600s, as the spectra are normalized to counts/hour.}, 
and $\lambda_{c}$ is the central wavelength of each pixel in \AA.

The accuracy and precision of the throughput correction was checked in the following way. 
For each high quality target galaxy at the redshift of the supercluster, the spectrum was multiplied 
by a fit to the Sloan Digital Sky Survey (SDSS) $i\arcmin$ filter curve using a quadratic interpolation to match the wavelength grid 
of each DEIMOS spectrum. Targets were chosen because they were
centered widthwise on the slit (serendipitous detections could fall anywhere on the slit) and supercluster members were chosen because
the range of half-light radii was well determined from the ACS imaging.

\begin{figure}
\plotone{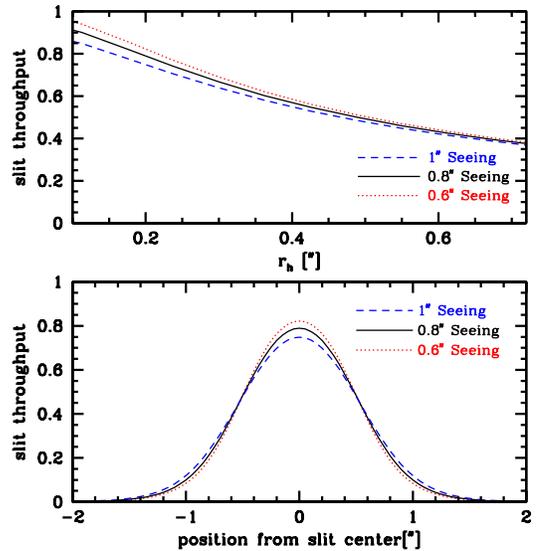}
\caption{Slit throughput ($\omega$) plotted as a function of half light radius (r$_{h}$) for a variety of different 
seeing conditions assuming the object is placed at the center of the 1$\arcsec$ slit (top panel), representing an absolute lower limit
on the amount of flux that must be lost by any galaxy when observed with our spectral setup. The lower panel plots the loss of flux as a function
of position along the minor axis (perpendicular to the spatial axis) of the slit for a galaxy with a half light radius of large LAEs (r$_{h}$=0.2$\arcsec$) 
for a variety of different seeing conditions. While the slit throughput has a moderately weak dependence on seeing and half light 
radius, the dependence on slit position is strong, falling off steeply when the object's position is more than 
0.4$\arcsec$ from the central position of the slit.}
\label{fig:seeing}
\end{figure}

A simulation was run in order to account for losses of light due to 
the finite spatial extent of the slit. Galaxies were simulated with exponential disk luminosity profile, 
half-light radii ranging from 0.34$\arcsec$ to 0.6$\arcsec$, based on values measured from ACS 
F814W data. For each simulated galaxy, the light profile was 
convolved with a Gaussian of FWHM comparable to the average seeing conditions under which 
our data were taken (0.9$\arcsec$). A slit of width 1$\arcsec$ and length 6$\arcsec$ was then placed on the galaxy, with 
the central part of the galaxy coincident with the central location of the slit. The total flux inside 
the slit was calculated for each simulated galaxy, with the slit throughput defined as the ratio of 
this quantity to the total flux in the absence of a slit. This slit throughput is plotted as a function of 
half light radius ($r_{h}$) and seeing in Figure \ref{fig:seeing}. In addition, a similar simulation 
was run to determine the slit throughput as a function of position from the slit center under a variety of different 
seeing conditions. Since we are most interested in this effect for LAE galaxies, an object with $r_{h} = 0.2\arcsec$ 
was used in the simulation, representing a reasonable limit to the sizes of large LAEs (see Overzier et al. 2006 or Venemans et al 2005).

Despite the functional dependence of slit-loss on the object's half light radius, the dependence is 
not particularly steep. For objects with $r_{h}$ $\leq$ 0.4$\arcsec$ the 
dependence is essentially linear. Thus, an average 
slit loss (1 - slit throughput) of 0.4 was adopted to correct each spectrum. Adopting an average slit loss correction was essential for the significant portion of DEIMOS objects 
which fall outside the coverage of the ACS mosaic and have no reliable half light radius measurements. 

\begin{figure}
\plotone{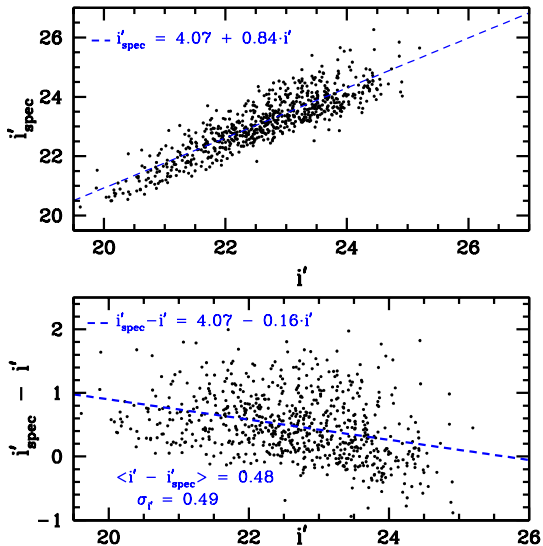}
\caption{LFC $i\arcmin$ magnitudes plotted as a function of magnitudes 
derived from the flux calibrated DEIMOS spectra (top panel) and as the difference between the 
spectral and LFC magnitudes (bottom panel) for all high quality (Q$\geq$3) spectra in the Cl1604 
supercluster. The best-fit relations are overplotted. The systematic offset between the spectral 
and LFC magnitudes in the brighter end of the lower panel is most likely due to an underestimation 
of slit losses or non-photometric considerations. This is not problematic as the offset drops to zero 
at the fainter end of the plot, the region that our LAE candidates populate. The large rms in the 
lower panel represent real uncertainties in flux calibration of the data, a trend that does 
not improve with decreasing brightness.}
\label{fig:SDSS}
\end{figure}

The flux density observed in the $i\arcmin$ bandpass for each spectrum is: 

\begin{equation}
\overline{f_{\lambda}}= \frac{\sum_{i=0}^n f_{\lambda,i} S_{\lambda,i} \delta_{\lambda,i} \lambda_{c,i}} {c \sum_{i=0}^n \frac {S_{\lambda,i} \delta_{\lambda,i}} {\lambda_{c,i}},}
\label{eqn:flamtot}
\end{equation}

\noindent where the sum is over the $n$ DEIMOS pixels that fall within the $i\arcmin$ bandpass and $S_{\lambda,i}$ is the $i\arcmin$ transmission as a function of wavelength.
The AB magnitude of each spectrum in the $i\arcmin$ band was then calculated by: 

\begin{equation}
i\arcmin_{AB,spec} = -2.5log(\overline{f_{\lambda}})-48.60 - \gamma \cdot sec(z),
\label{eqn:ABmag}
\end{equation}

\noindent with $\gamma$ being the airmass term for Mauna 
Kea\footnote{http://www.cfht.hawaii.edu/Instruments/ObservatoryManual/ CFHT\_ObservatoryManual\_(Sec\_2).html}. 
This spectral $i\arcmin_{spec}$ magnitude was then compared to our Palomar Large Format Camera (LFC; Simcoe et al. 2000) photometry 
(see G08 for details). Since the 
slit positions were determined from the LFC imaging, there were cases where there were 
noticeable ($>$ 1$\arcsec$) positional errors. Thus, galaxies not centered or absent from the slit 
or those with photometric flags were removed from the sample. The derived spectral magnitudes of the remaining galaxies are 
plotted against the LFC photometric magnitudes in Figure \ref{fig:SDSS}.

The rms scatter of the spectral magnitudes between $i\arcmin$ = 19.5 and $i\arcmin$ = 25 is 0.49 magnitudes, corresponding 
to an $\sim$ 60\% uncertainty in any absolute flux 
measurement. While the range of magnitudes are brighter than the average magnitude (or magnitude limit) 
of the LAE candidates in our sample, we adopt this rms as being reflective of the uncertainty in Ly$\alpha$ 
line flux measurements. In addition, the spectral magnitudes are systematically fainter on 
average by 0.48 magnitudes (lower panel of Figure \ref{fig:SDSS}). While this offset also corresponds to a 
bias of $\sim$ 60\% for absolute flux measurements, this is less of a concern than the rms scatter 
for several reasons. First, the trend in the systematic offset as a function of 
magnitude tends towards zero at fainter magnitudes. If a magnitude-size relation is assumed for our target 
galaxies, the observed trend suggests that any offset comes from underestimating slit losses 
for the brighter target galaxies. Since we are interested in Ly$\alpha$ line fluxes, emission which 
originates from host galaxies that have typical $i\arcmin$ magnitudes fainter than our dimmest target galaxy (i.e., $i\arcmin$ 
$>$ 25.2), this systematic will not adversely affect our measurements. 
While we use a slit throughput of 0.8 (see the following section) when calculating the line fluxes for the purposes 
of deriving LAE properties such as EWs or SFRs,  
a full slit loss simulation used in calculating the luminosity function is undertaken in Section 5.4.
Finally, we approach all measurements from the bottom---,i.e., 
erring on the side of underestimating the true flux of the galaxies so that our 
measurements will be a strict lower limit to compare with other surveys. We therefore ignore this systematic and include only the rms error when calculating line fluxes. 

\subsection{Line Flux Measurements}
For each single emission line galaxy, the one-dimensional spectrum was inspected, and three bandpasses were chosen to measure the emission line flux. The first bandpass 
encompasses the entirety of the emission line, avoiding any instrumental or reduction artifacts. 
The other two bandpasses were chosen to be relatively sky line free regions blueward and redward of the emission line, as close 
to the emission line in the dispersion dimension as the data would allow, set to a minimal width of 20 \AA. A linear model was fit to each spectrum in 
the blueward and redward bandpasses to mimic the continuum throughput. The model parameters were fit with a $\chi^{2}$ 
minimization routine, with the associated errors calculated from the covariance matrix. While a continuum fit was 
typically unnecessary for LAE objects, as the associated background was formally consistent with zero in 
most cases, the above procedure was implemented to accurately measure the line flux of low-$z$ single-emission line galaxies 
used as a comparison (see Section 3). 

The resulting model background was subtracted 
from each spectrum in the emission line bandpass, with the total flux in each bandpass measured by: 

\begin{equation}
F_{Ly\alpha} [\rm{\frac{ergs}{cm^{2} \, s}}] = \sum_{i=0}^n \left( f_{\lambda,i} \, \delta\lambda_{i} - B(\lambda_{i}) \right) \, \frac{1}{\omega_{slit},}
\label{eqn:ftotemission}
\end{equation}

\noindent where $B(\lambda_{i})$ is the model at each wavelength, $\delta\lambda_{i}$ is the size of the pixel at each wavelength, $\omega_{slit}$ is the slit 
throughput, and $f_{\lambda,i}$ is defined in Equation \ref{eqn:flam}. The slit throughput used in the calculation of the Ly$\alpha$ line fluxes was set to 0.8, 
appropriate for a target galaxy with a half-light radius of 0.2$\arcsec$ in 
0.9$\arcsec$ seeing. As most LAE candidates are not in the 
middle of the slit (as a target would be) and since the slit-throughput function 
remains below 80\% for galaxies centered on the slit for all but the smallest half-light radii 
($r_{h} \la 0.1\arcsec$), the flux measured in this way still represents a lower limit to the true 
flux coming from the galaxy.
Tables \ref{tab:nondettable} and \ref{tab:dettable} list the name, 
redshift (assuming the line is Ly$\alpha$), right ascension and declination (assuming the 
serendip is at the center of the slit widthwise), the confidence class, line flux (minimally corrected for flux losses 
due to the slit as in the above equation), line luminosity, measured or 3$\sigma$ limiting magnitudes,
the EW of the Ly$\alpha$ line, and the observed wavelength of each 
LAE candidate.

The associated errors for each flux measurement were derived from a combination of (response corrected) Poisson errors from 
each spectrum and the errors associated with the background model, as well as 
the flux calibration error discussed in the previous section. There can also be significant 
systematic errors associated with the bandpass choices. Limiting the size of the emission line bandpass can significantly 
underestimate the true line flux, while an overextension of the limits can introduce significant noise into 
the measurement. A select group of galaxies, spanning the dynamic range of the spectra measured in this manner, were analyzed 
in order to estimate the magnitude of this error. In all cases 
the systematic errors derived for a ``reasonable" range of bandpass choices were completely dwarfed by Poisson errors. 

\subsection{Flux Limit and Spectral Completeness}

Since our search depended almost entirely on human detection of sources, accurately quantifying the completeness limit of the objects 
detected is more difficult than in searches that use automatic peak finding algorithms. The human eye, while being very 
good at discriminating between spurious and real detections and at finding irregularities in data (serendips in 
our case), is subject to a variety of effects which are difficult to quantify. To roughly quantify our completeness limit we simulated one hundred slits, each 55 by 8192 pixels corresponding to 6.5$\arcsec$ by 2700 \AA\ at the 
DEIMOS 1200 l mm$^{-1}$ grating plate scale. These data were first simulated using the noise and background properties measured from 
actual DEIMOS two-dimensional spectra in regions where features and poor-sky subtraction were absent. These feature-free, 
artifact-free regions were collapsed into one-dimensional spectra using the same method used by \emph{spec2d} in extracting one-dimensional 
spectra of target galaxies. Each of the two-dimensional spectra were populated with flux values that mimicked the properties 
of the real two-dimensional spectra, creating in essence one-hundred 6.5$\arcsec$ ``blank-sky" slits. These simulated blank-sky slits 
were populated with objects that varied in both intensity and frequency. For each simulated slit, 
between zero and four objects were placed on the slit, characterized by two-dimensional Gaussians 
with freely varying amplitudes, dispersions in both the spatial and spectral dimensions, spatial locations, and central wavelengths. 
Noise was also introduced to each Gaussian to properly simulate the counting error associated with observing actual galaxies. 
The slits were populated so that a slit had zero objects 50\% of the time and between one and four objects 50\% of the time. 
In addition, the heights and dispersions of the Gaussians were constrained so that the objects would have reasonable 
flux values, i.e. values corresponding to an order of magnitude both fainter and brighter than the faintest and brightest 
single-emission line object detected in our data. 

Each of the two-dimensional slits was then analyzed by one of the authors (BL) in blind observations using a fashion similar to that used for the original data. 
The conditions that were present when observing the original slits were re-created to the best of our ability
(e.g., the time spent on each slit, the method of looking for detections, the software used). For every simulated object detected in the two-dimensional 
slits, a one-dimensional spectrum was created using methods similar to \emph{spec2d}. A catalog of generated objects was compared to the catalog of objects detected by eye and the remaining objects that went undetected 
in the data were then similarly extracted. If we set the completeness limit at the faintest object detected nearly 100\% of the 
time, this limit corresponds to objects with significances between 105$\sigma$ and 111$\sigma$ in the 
two-dimensional data, or a one-dimensional significance of 7$\sigma$. This significance translates to a 
completeness limit of $1.9\times10^{-18}$ ergs s$^{-1}$ cm$^{-2}$ for a 7200s exposure time, decreasing slightly for our masks with longer integration times. This 
completeness limit is consistent with the line flux analysis done in Section 3.3 (see Figure \ref{fig:fluxcomp} and 
associated discussion), suggesting that this limit is close to the actual completeness limit of the survey.  

\section{Emission Line Tests}

The large spectral coverage and moderately 
high resolution of DEIMOS give us a distinct advantage over narrowband imaging searches for LAEs
or searches with small spectral coverage, as we are able to differentiate the 
Ly$\alpha$ line from other emission lines that are typically confused for it. 
The lines which are the most prevalent contaminants in searches for Ly$\alpha$ 
emission are the 3727 \AA\  [OII] doublet, [OIII] at 5007 \AA, H$\beta$ at 4861 \AA, or H$\alpha$ at 6563 \AA. 

The most insidious contaminant in many LAE surveys is the [OII] doublet 
(rest frame separation 2.8\AA). For our spectral setup this line would be 
observed at a redshift of 0.71 $\leq$ $z$ $\leq$ 1.41 and is usually resolved with 
the 1200 l mm$^{-1}$ grating. A small fraction of the [OII] doublets are 
unresolved due to a combination of galactic rotational effects and the 
slit being oriented along the major axis of the galaxy. In this case the [OII] line can still be 
discriminated from Ly$\alpha$ by the asymmetry of the line. 
The nebular Ly$\alpha$ line is typically characterized by its strong asymmetry, 
with suppression of line flux in the blueward portion and, 
in some cases, an extended redward tail. A blended (unresolved) [OII] line in normal star forming 
regions (in the absence of an active galactic nucleus (AGN)) exhibits
asymmetry opposite that of Ly$\alpha$, with an extended 
tail in the blueward portion of the line (Osterbrock 1989; Dawson et al. 2007).  
Galaxies emitting H$\alpha$, H$\beta$, or [OIII], in cases of even moderate S/N, can be easily 
distinguished from Ly$\alpha$ by other associated spectral features. The 5007 \AA\ [OIII] 
line is typically seen with 4959 \AA\ [OIII] and 4861 \AA\ H$\beta$ with varying degrees of relative
intensities (Baldwin et al. 1981). The 6563 \AA\ H$\alpha$ line can be identified by 
two accompanying SII lines at 6716 \AA\ and 6730 \AA\ and two [NII] lines at 6548 \AA\ and 6583 \AA,
also with varying degrees of relative intensity.
Many spectra originally classified as single-emission line objects were recognized as low 
redshift interlopers through the identification of faint associated lines. 

\begin{figure}
\plotone{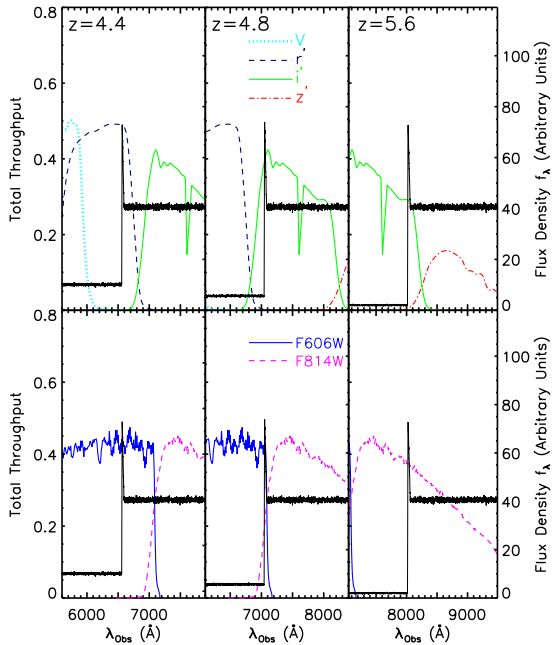}
\caption{Synthetic Ly$\alpha$ spectra at three different redshifts spanning the range of redshifts of our sample overlaid on the 
four ground-based filter transmission curves (top three panels) and two ACS filter transmission curves (bottom three panels) for which we have coverage. 
The continuum break over the Ly$\alpha$ line is modeled based on narrowband imaging measurements of the continuum break (H04). For 
lower redshift LAEs, the only band able to detect the continuum break is the ground-based V band. At
the redshift of most of our sample ($z \sim 4.8$), the Ly$\alpha$ line just passes the coverage of the ground-based $r\arcmin$ band 
and lies just at the red end of the F606W band. At higher redshifts the Ly$\alpha$ line is comfortably situated in the F814W band 
and at the red end of the ground-based $i\arcmin$ band, giving us significantly more power to discriminate between interlopers and genuine
LAEs at these redshifts. The throughput of the V band is scaled down and the throughput of the $z\arcmin$ band is scaled up for clarity.}
\label{fig:LYAbands}
\end{figure}

For the remaining 39 objects that were classified as genuine single-emission 
line objects, several tests were performed to further remove any low redshift 
interlopers. The Ly$\alpha$ line is characterized by a large 1.3--4.5 mag continuum break blueward of the line due to 
attenuation of Ly$\alpha$ photons by intervening neutral hydrogen (H04). Initially, the 
spectral data were inspected, and 10 single-emission line serendip exhibiting 
appreciable continuum blueward of the emission feature relative to any redward 
continuum was eliminated as a potential LAE candidates. The imaging data were 
also useful in discriminating single-emission line serendips in this regard, as 
the photometric filter setup would also, in many cases, 
probe the continuum break across the Ly$\alpha$ line (see Figure \ref{fig:LYAbands}). Each single-emission line 
serendip detected in one or more of the photometric bands was required 
to exhibit a continuum break over filters blueward and redward of the line. 
Since most of these objects are extremely faint in the imaging (if they are 
detected at all), requiring a strong continuum break over the emission line is, 
in almost all cases, similar to requiring that the object drop out of any band 
blueward of the emission line. Our bluest LFC and ACS bands, $r\arcmin$ and F606W, are 
situated so that either would pick up a significant amount of continuum flux 
from any LAE at the bluer end of our detection limit ($\lambda$ $\leq$ 7000 \AA, $z_{Ly\alpha}$ $\leq$ 4.75). 
For objects such as this we had to rely on Subaru Suprime-cam V-band data 
to discriminate between potential LAEs and low-$z$ interlopers. Any galaxy detected 
in the V-band data was excluded as an LAE candidate due to the relative shallowness 
of the image (see Section 4.1 for details on the depth of the photometry). All but two of 22 single-emission line
galaxies that were eliminated as potential LAE candidates through the above tests failed the continuum break test. 
The two single-emission line low-$z$ interlopers that did not fail this test were among the 10 galaxies 
that failed the spectral continuum break test. In addition, each single-emission line serendip that was detected in the photometric data was also 
inspected visually, and any objects with large ($>$ 2$\arcsec$) angular extents were 
classified as low-$z$ interlopers. Six of the 22 single-emission line low-$z$ interlopers were rejected by this test,
although in all cases these galaxies had failed at least one of the two previous tests.

\begin{figure*}
\plotone{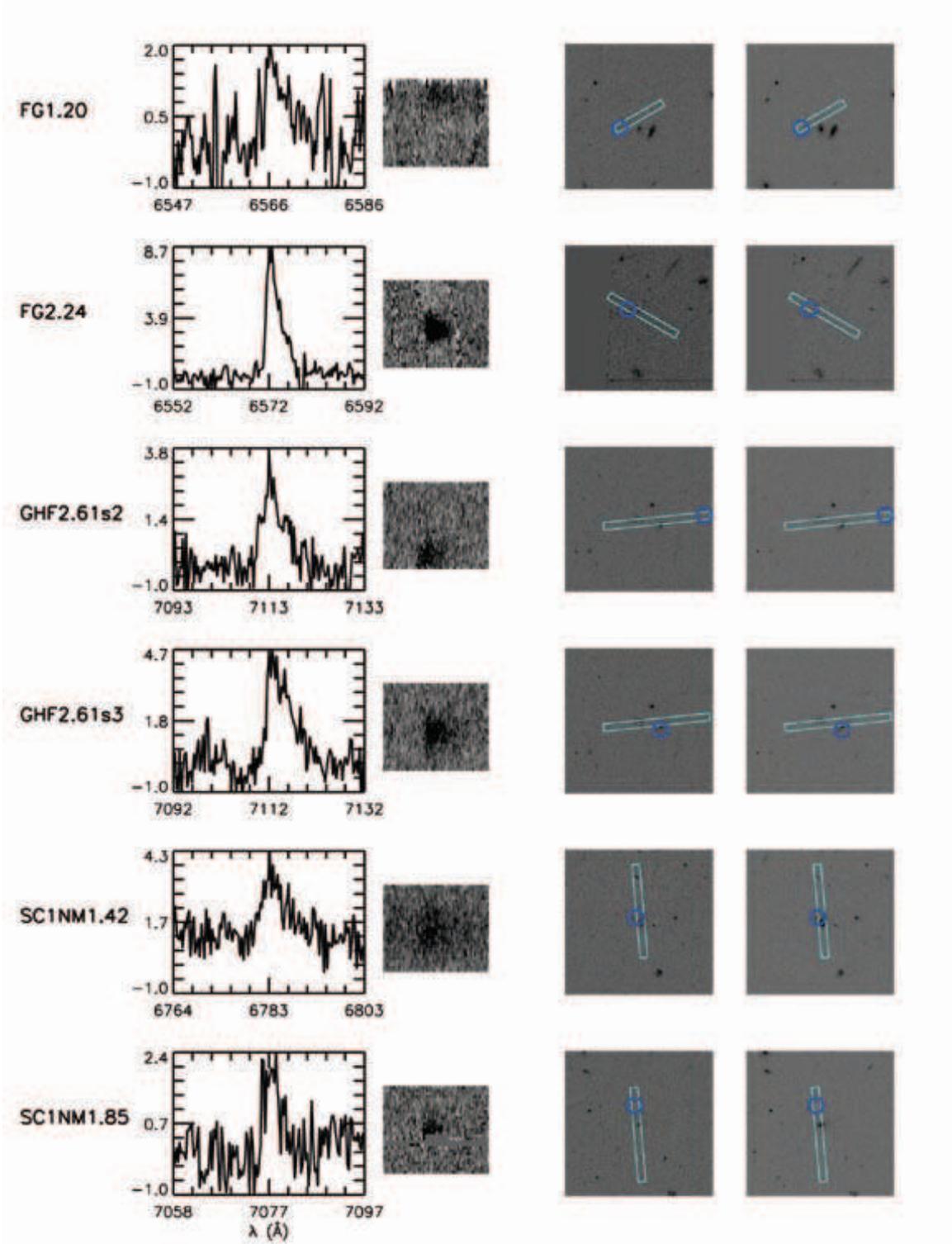}[p]
\caption{Spectral ID, cutout of flux calibrated DEIMOS one-dimensional spectrum uncorrected for slit
losses in units of $\mu$Jy, and cutout of DEIMOS two-dimensional
spectrum for each LAE candidate. Postage stamps of the ACS F606W and F814W images (when available)
or the LFC $r\arcmin$, $i\arcmin$, and $z\arcmin$ show the DEIMOS slit (box) and the LAE candidate
position (circle) either from the detected position or inferred assuming the LAE fell in the
middle of the slit (widthwise). LAE candidates 1 through 6.} 
\label{fig:mosaic1}
\end{figure*}

\begin{figure*}[p]
\plotone{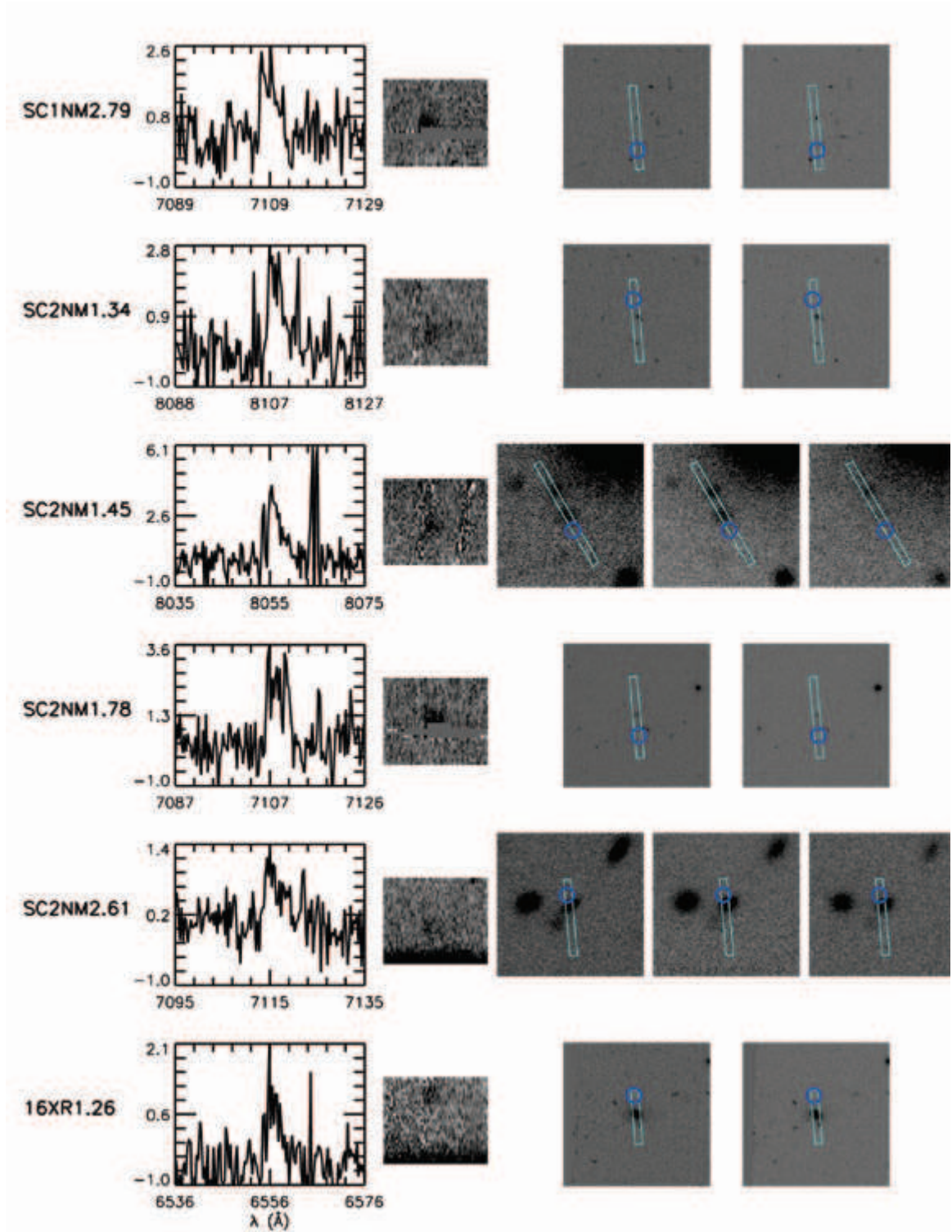}
\caption{LAE candidates 7 through 12.}
\label{fig:mosaic2}
\end{figure*}

\begin{figure*}[p]
\plotone{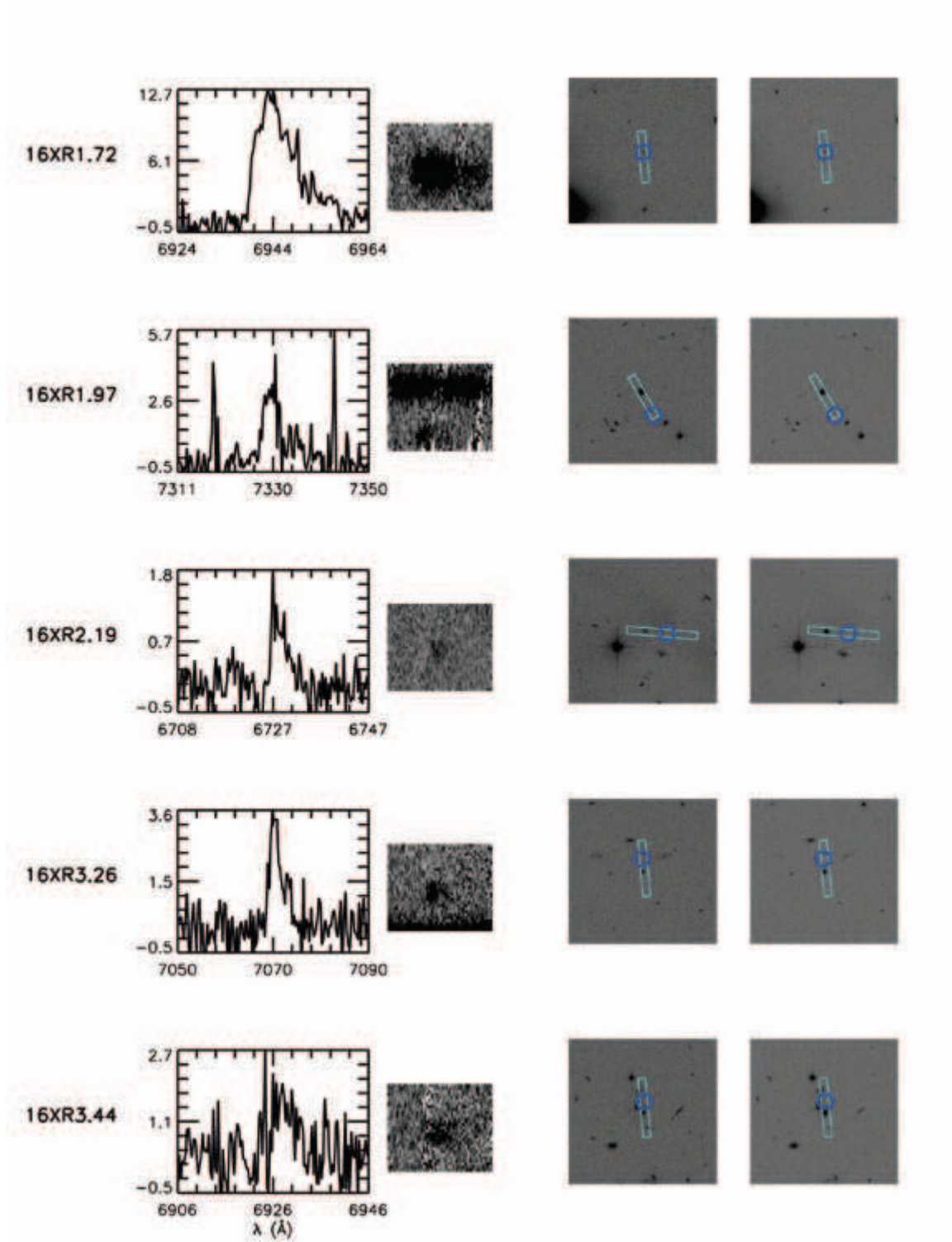}
\caption{ LAE candidates 13 through 17.}
\label{fig:mosaic3}
\end{figure*}

Of the original 39 single-emission line cases, 17 objects survived the 
previous tests. A small subset of these objects were insensitive to 
these tests, as the single-emission line object was superimposed spatially in 
the spectral data with either the target or another serendip. In such cases, the 
single-emission lines were checked against a variety of nebular emission lines 
at the redshift of the superimposed target or serendip to verify that it could 
not simply be an unusual emission feature coming from the same galaxy. In many 
cases, however, it was clear from the morphology or positions of the lines that the two emission features  
originated from two separate sources. In some cases the two superimposed 
objects were resolved in the ACS data, and the continuum break 
test was used on one or both of the galaxies, depending on whether the identity of 
the single-emission line source was certain. More frequently, however, the two 
objects remained unresolved in the \emph{HST} data, so we  
include them in our sample. Of the 17 objects that survived the original single-emission line tests,
all 17 passed the tests described above. These 17 objects comprise our LAE sample (see Figures \ref{fig:mosaic1}-\ref{fig:mosaic3}). 

\subsection{Line asymmetry}

\begin{figure}
\plotone{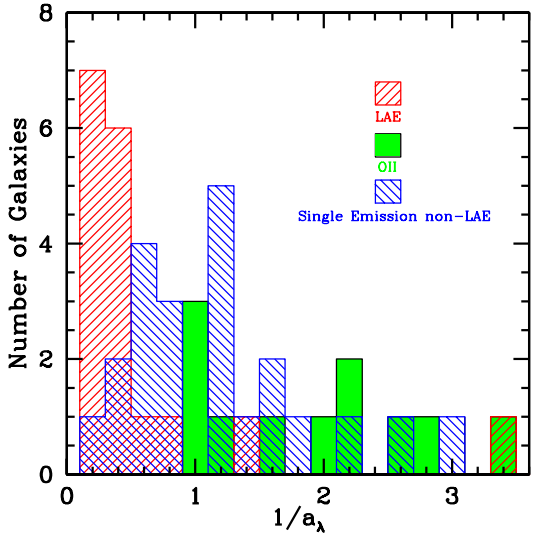}
\caption{The inverse of the asymmetry parameter measured for the emission lines of three different populations: known blended [OII] emitters 
($0.7<z<1.4$), low-$z$ single-emission line galaxies, and our 17 LAE candidates. This parameter easily discriminates even blended [OII] 
emission lines, which are strongly skewed toward high values of the inverse of the asymmetry parameter, from LAE lines which typically exhibit values much 
less than unity. The low-$z$ single-emission line galaxies are less easily discriminated by this test, with a distribution that is 
centered around symmetric (1/a$_{\lambda}$=1) line profiles and wings that extend into the phase space of the other two populations. All 
LAE candidates with higher inverse asymmetries (1/$a_{\lambda}>0.75$), including one very high value, are low-quality candidates.}
\label{fig:asymmetry}
\end{figure}

Another discriminator used on the individual single-emission line spectra was a 
computation of the wavelength asymmetry parameter (Dawson et al.\ 2007). 
Briefly, the asymmetry parameter, a$_{\lambda}$, is defined as: 

\begin{equation}
\frac{1}{a_{\lambda}}= \frac{\lambda_{c} - \lambda_{10,b}}{\lambda_{10,r} - \lambda_{c},} 
\label{eqn:asymm}
\end{equation}

\noindent where $\lambda_{c}$ is the central wavelength of the emission, defined as the point of 
maximal flux in the line profile, and $\lambda_{10,r}$ and $\lambda_{10,b}$ 
are the wavelengths where 
the flux first exceeds 10\% of the peak flux redward and blueward of the line, 
respectively. This diagnostic can be used to further  
discriminate single-emission lines that exhibit standard Gaussian (Voigt) 
profiles such as H$\beta$, H$\alpha$, [OIII] (1/$a_{\lambda}$ $\sim$ 1), or 
a blended 3727 \AA\  [OII] doublet (1/$a_{\lambda}$ $>$ 1) from 
a higher redshift Ly$\alpha$ line that exhibits strong asymmetry in the opposite direction.
While this test can be a useful diagnostic in a 
statistical sense, an asymmetry parameter of $1/a_{\lambda}$ $\geq$ 1 was not a 
strong enough constraint to rule out an object as an LAE candidate if 
it had passed all the previous tests. This is because several processes (instrumental broadening, local underdensities
of HI regions, etc.) can cause the LAE emission to appear symmetric. 
Conversely, an object which had 
failed one or more of the above tests was not reclassified as a potential LAE 
based on an unusually high (1/$a_{\lambda}$ $<$ 1) asymmetry parameter, as low redshift lines can, 
under rare circumstances, exhibit strong redward-skewed asymmetry (see for example object D21 in MS04). Therefore, this diagnostic 
was used only to discriminate between high quality LAE candidates and poorer 
quality candidates, rather than distinguishing genuine LAEs from interlopers. Figure 
\ref{fig:asymmetry} shows a histogram of the inverse of the asymmetry parameter of the known lower redshift 
single-emission line objects, a population of blended [OII] emitters (confirmed by 
other associated lines present in the spectrum), and our 17 LAE candidates. The objects clearly separate out; the LAE
candidates primarily occupy the high asymmetry (low inverse asymmetry) portion of phase space, the low-$z$ interlopers 
are distributed around unity (symmetric), and the [OII] galaxies are primarily situated in the region of phase space opposite that of the LAE
candidates. In fact, all but three LAE candidates (all Quality 1; see below) have inverse asymmetry parameters less than 0.75. 

\subsection{LAE Confidence Classes}

Each of the 17 LAE candidates was assigned a quality class. Quality classes are assigned to LAE candidates in a fashion nearly identical to that of S08 and are defined as 
follows: Quality 1 objects pass all of the above tests, 
but show no additional indicators of being genuine LAEs. Objects which are 
Quality 1 do not exhibit any asymmetry (or exhibiting blueward-skewed asymmetry) in 
their line profiles and are non-detections in all photometric bands. These 
objects are our least secure 
candidates, nearly equally likely to be low luminosity foreground galaxies as 
LAEs. Quality 2 and 3 objects all similarly pass the interloper tests but also
show strong asymmetric line profiles. A few of these objects are 
detected in one or more photometric bands, further increasing our confidence in 
these objects as genuine LAEs, but it is the asymmetric line profile which is 
the defining characteristic of the higher confidence classes. Both Quality 2 and 
Quality 3 candidates represent our highest level of confidence that an object is 
a genuine LAE. However, Quality 2 objects are superimposed with a target 
or another serendip spatially on the slit. Thus, the flux measurements of the Quality 
2 objects could be significantly dimmed by extinction from the interstellar medium (ISM) of the 
foreground galaxy or boosted through galaxy-galaxy lensing. This additional, unknown component of the uncertainty makes it necessary to 
exclude Quality 2 galaxies from certain parts of the analysis.

\subsection{Flux and Redshift Tests}

The tests discussed in the beginning of Section 3 can 
only be used to rule out objects as LAEs, not to prove that any 
particular object is definitively an LAE. The tests in the following two sections
explore the statistical similarities or differences between 
LAE candidates and the low-$z$ interlopers, giving us further confidence that the 
LAE candidates represent a unique and separate population. 

\subsubsection{Effective Redshift Test}
First we compare the observed wavelengths of the 
single-emission lines in the low-$z$ interloper population to the observed 
wavelengths of the Ly$\alpha$ lines in the LAE candidates. The low-$z$ single-emission 
line interlopers are comprised of some combination of [OII], H$\beta$, [OIII], 
and H$\alpha$ emitters and therefore cannot be given definite redshifts. Following the analysis
done in S08, we have recast the low-$z$ interlopers in terms 
of an effective redshift: the redshift that the object would have if the line were Ly$\alpha$, 
such that $z_{eff, Ly\alpha}$ = ($\lambda_{em}$/1215.7 - 1). 

The idea of this test is that the low-$z$ single-emission line interlopers, if 
they truly are comprised of a mix of the aforementioned lines, should be, in the 
absence of any instrumental effects, equally distributed in effective redshift 
(wavelength) space. An object at a redshift of $z$ = 0.35 emits the 5007 \AA [OIII] 
line at $\lambda_{obs}$ = 6759 \AA\ and 6563 \AA\  H$\alpha$ at $\lambda_{obs}$ = 8860 \AA, both of which could mimic 
single emission lines under a variety of different conditions. These effects could be 
(1) instrumental: the placement of the slit on the slit mask truncating either 
the blue or red end of the CCD response; (2) atmospheric: a bright night sky line 
masking the second emission line; or (3) a result of galactic processes: a low 
level AGN which exhibits strong [OIII] emission but little to no Balmer emission, 
or a starburst galaxy having strongly suppressed forbidden transitions relative 
to the strength of the Balmer lines. In any of these cases, the chance is more 
or less equal that the single-emission line galaxy will show up as the blue or 
the red emission line. The redshift distribution of the LAE population should be 
strongly biased towards the lowest redshifts to which we are sensitive, as we probe 
successively shallower in the luminosity function as the LAEs move to higher 
redshifts. Thus, if the LAE population represents a truly different population 
than the low-$z$ single-emission line interlopers, the redshift histograms 
should differ significantly. 

\begin{figure}
\plotone{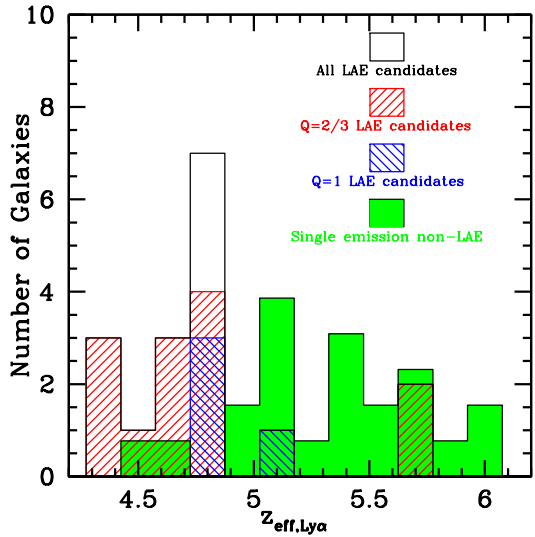}

\caption{Redshift histogram of the high quality (Q=2,3) and low quality (Q=1) LAE candidates. A strong peak 
can be seen at $z\sim4.8$ as well as less pronounced peaks at $z\sim4.4$ and $z\sim5.7$. Conversely, the 
overplotted low-$z$ interloper population (plotted in terms of $z_{eff,Ly\alpha}$; see Section 3.3.1) is distributed nearly
symmetrically around $z\sim5.1$, with a much slower falloff at high redshifts.}
\label{fig:zhist}
\end{figure}

Figure \ref{fig:zhist} shows the comparison in effective redshift space between 
the 22 low-$z$ interlopers, the 4 Q=1 and the 13 Q=2,3 LAE candidates. The 
low-$z$ single-emission line interlopers are more or less evenly distributed 
across $z_{eff, Ly\alpha}$ with two important exceptions. There are no interlopers 
shortward of  $z_{eff, Ly\alpha}$ = 4.4, possibly due to the prevalence of H$\alpha$ as the 
unknown single emission line in the interloper population. The rest wavelength 
of the H$\alpha$ line has $z_{eff, Ly\alpha}$  = 4.398 so if the interloper population does 
consist primarily of $H\alpha$ emitters, few galaxies would be seen blueward of this 
limit. Another reason for this drop-off in detections could be 
the significant drop in DEIMOS sensitivity blueward of $\sim$ 6600 \AA\  for our spectral 
setup. The second drop-off in detections occurs at  $z_{eff, Ly\alpha}$ $>$ 
6.1, most likely due to the significant decrease in DEIMOS sensitivity and the decrease in 
significant sky line free spectral windows redward of $\sim$ 8700 \AA\ (see Figure \ref{fig:through}).

The LAE population is strongly peaked towards the low 
end of our redshift sensitivity. A very noticeable peak exists at $z \sim 4.8$, which 
may represent a real clustering of the LAE population in projection space or 
could simply be an artifact of the sensitivity issues discussed in the previous 
paragraph, as the DEIMOS sensitivity peaks at $\sim$ 7000 \AA\  for our setup. More likely, it is some combination of these two effects 
(see Section 5.2 for a discussion). The Q=1 LAE candidates, which are our least secure 
candidates, are surprisingly consistent with our higher confidence Q=2,3 
population, also peaking around $z_{Ly\alpha}$ $\sim$ 4.8. There are two Q=2,3 
candidates at $z \sim 5.7$ which are unexpected, given our prediction that the 
LAE population should be strongly peaked towards the low-redshift end. 

\subsubsection{Line Flux Test}
The second of these tests explores the possibility that the LAE candidate population represents a
lower luminosity subset of the single-emission line interlopers. A majority of single-emission 
line interlopers were ruled out by broadband detections, i.e., not exhibiting a sufficiently strong 
continuum break over the feature to be plausibly identified as Ly$\alpha$. All of the single-emission 
line interlopers were detected in the 
photometry. Conversely, the majority of the LAE candidate population were not detected in any 
of the three LFC bands nor the two ACS bands. Thus, the LAE candidate population 
clearly represents a class of objects that are significantly dimmer in continuum luminosity. If 
the LAEs are truly drawn from the same population as the low-$z$ interlopers, their line 
luminosities should similarly scale down. This test provides a quantitative 
statistical tool to differentiate the LAE candidates from the lower luminosity tail of the 
single-emission line interloper luminosity function. This test is not sensitive to the case where the LAE 
candidates represent a population of dwarf starbursting galaxies with higher line 
luminosity relative to their continuum brightness (Fricke et al. 2001; Guseva et al. 2003;
Kehrig et al. 2004; Izotov et al. 2006).

\begin{figure}
\plotone{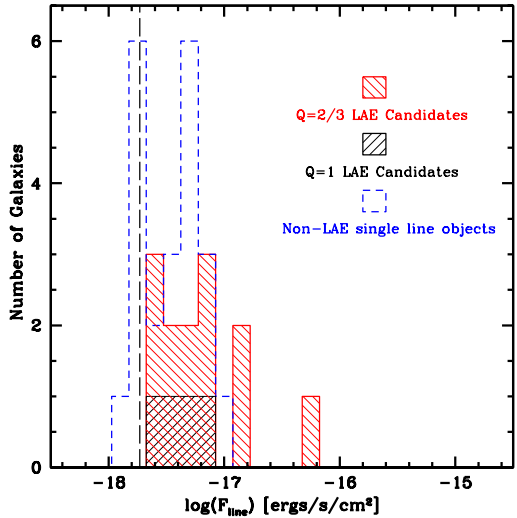}
\caption{Histogram of line fluxes uncorrected for slit losses of our LAE candidates and confirmed low-$z$ single-emission
line objects. While the two distributions overlap, the low-$z$ interloper population is characteristically dimmer than both the high (Q=2,3) and 
low (Q=1) quality LAE candidates. The difference in the observed distribution suggests that the LAE candidates are not primarily comprised of low
luminosity single-emission line objects at low ($z < 1.4$) redshifts. The vertical long dashed line 
represents our adopted completeness limit of $1.9\times10^{-18}$ ergs s$^{-1}$ cm$^{-2}$}
\label{fig:fluxcomp}
\end{figure}

Figure \ref{fig:fluxcomp} shows the comparison between the line fluxes of the single-emission line 
interlopers relative to the LAE candidates. The Q=2,3 LAE candidates are, on average, brighter 
than the single-emission line interloper objects, with the mean line flux about 0.5 dex higher than 
the interloper population. The average magnitude of the interloper population in the band which 
best samples the continuum emission near the emission feature is 23.5 mags. In contrast we can adopt 
the LFC $i\arcmin$ 3$\sigma$ limit of 24.3 mags as the upper bound on the continuum flux of LAEs 
that are not detected in the photometry (a conservative limit as many of the candidates are undetected in
the ACS images which have a 3$\sigma$ depth of $\sim26$ mags). This limit on the continuum 
flux requires the LAE candidates, if they are instead low-luminosity, low-$z$ interlopers, to have  
line equivalent widths (EWs) at least 10 times greater than the average EW of the known single-emission 
line interlopers (5.4 \AA). Such high EWs are certainly plausible in dwarf galaxies undergoing a 
starbursting event where the EWs of H$\alpha$ (usually the strongest lines in optical starbursting spectra) are the range 50-150 \AA\  
(Kennicutt 1998, Petrosian et al. 2002) and have been measured as high as $\sim$1500\AA\ (Kniazev et al. 2004, Reverte et al. 2007). However, 
such objects are uncommon, and we would expect to observe other associated lines (e.g., [NII], H$\beta$, [OIII]) 
in the data, which we do not. It is interesting to 
note that if we adopt a standard ratio for log$[L([NII]$ 6585 \AA$)/L(H\alpha)]$ and log$[L([OIII]$ 5007 \AA$)/L(H\beta)]$ of -0.45 
for star-forming galaxies (Baldwin et al. 1981; Brinchmann et al. 2004; Shapley et al. 2005; Yan et al. 2006), the bulk of our LAE candidates ($\sim$60\%) are sufficiently brighter 
than the completeness limit so that the associated lines would be detected if the emission were instead H$\alpha$ or H$\beta$.

The Q=1 LAE candidates are essentially identical to the fluxes of the low-$z$ interloper 
population, with a mean flux of $4.8\times 10^{-18}$ ergs s$^{-1}$ cm$^{-2}$ as compared to the mean flux of the interloper 
population of $3.9\times 10^{18}$ ergs s$^{-1}$ cm$^{-2}$. While this similarity may be an indication that the Q=1 LAE candidates 
contain at least some low-$z$ interlopers mixed in with genuine LAEs, it also may be misleading. 
The average upper limit on the magnitude of the Q=1 candidates in the filter sampling the continuum 
surrounding the emission feature is 24.9, nearly 1.5 magnitudes dimmer than for the
single-emission line counterparts. While the line fluxes of these two populations are similar, the EW of the 
$Q$=1 LAE candidates would still necessarily have to be a factor of 4 higher than the 
interloper population. In addition, the Q=1 line fluxes fall near the completeness limit of 
$1.9\times10^{-18}$ ergs s$^{-1}$ cm$^{-2}$ and near the low-flux tail of the line flux measurements of the high quality 
(Q=2,3) LAE candidates. We would expect, independent of the redshift range, an 
inverse relationship between the number of detections and the line flux down to the completeness 
limit and a steep falloff in detections thereafter. If the Q=1 LAE candidates constitute real detections of genuine LAEs, this 
would be the behavior we observe in the data. Thus, it may be that these low quality candidates simply represent the fainter flux end of the LAE 
population, and their lower S/N prevents them from reliably being classified as high quality 
candidates. 

\subsection{Composite Spectra}

Previous tests focused on measurements of individual spectra of galaxy signatures at or near the
flux limit, making these measurements susceptible to noise effects. While we compare the ensemble properties 
of the galaxy populations, which is less sensitive to noise variations in the data than the comparison of individual 
measurements, an alternative is co-adding of the spectra in order to increase the S/N.  

\begin{figure}
\plotone{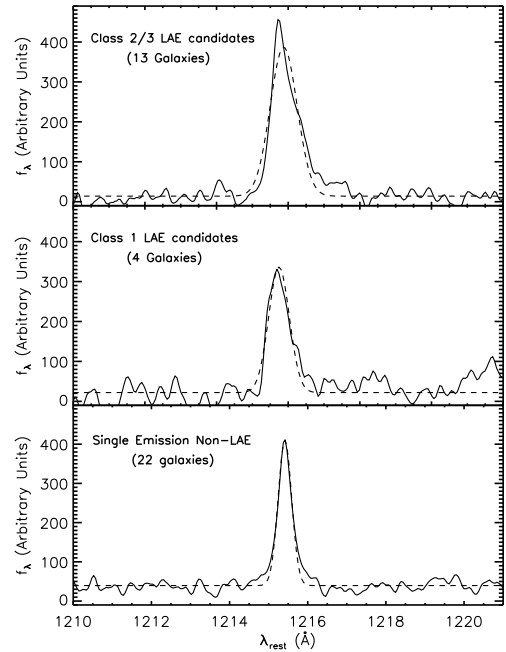} 
\caption{Composite spectrum of the LAE candidates and the known low-z interlopers 
smoothed with a Gaussian smoothing kernel of $\sigma$=1.5 pixels. Each co-addition is done using
luminosity weighting (see the text). The dashed lines show Gaussian fits to the line profiles. Both the 
high (Q=2,3) and low (Q=1) quality LAE candidates are poorly fit by a Gaussian model.} 
\label{fig:lineshapes}
\end{figure}

To properly retain the overall spectral properties of the constituent objects (e.g., line 
shapes, resolution, velocity dispersions, etc.) and to avoid averaging out faint features, it is 
necessary when coadding galaxies to determine the redshift as accurately or in as consistent a 
manner as possible. While we were able to determine redshifts for our 
interloper sample through the centroiding provided by \emph{spec2d}, the LAE population was 
problematic because of the uncertainty in determining the true peak of the line. In the 
absence of any other knowledge about the true profile of the Ly$\alpha$ emission in each galaxy 
(other than its asymmetry), the assumption was made that the wavelength associated with the 
peak flux in each emission line profile represented the central wavelength for that emission. Lack of knowledge about the 
shape of the true line profile introduces a significant ($\delta_{z} \sim 0.002$) absolute 
error in the redshift measurements. However, since this measurement is made in a consistent way for each 
spectrum, the relative error in the redshifts (the important quantity for coadding purposes) 
between any two spectra is quite small. Thus, any asymmetry in the original
line profiles should be preserved through this process. 
 
Each galaxy spectrum was then ``de-redshifted" to its rest frame, or, in the case of the single-emission line 
interlopers, the effective rest frame (see Section 3.3). Each 
rest frame spectrum was interpolated onto a pixel grid of common size, chosen to subsample the 
lowest (rest-frame) pixel scale. The resulting spectra were then added together in the following 
two ways: (1) each spectrum was normalized by the galaxy's total spectral flux (uniform 
weighting), or (2) galaxies were added together with no normalization (luminosity weighting). 
In both cases, the flux of each pixel in the co-added spectrum was populated using a Poissonian variance 
weighted mean of the pixel values at each wavelength in the individual spectra. 

Figure \ref{fig:lineshapes} shows the luminosity-weighted coadded spectrum for three different 
populations: the high quality (Q=2,3) LAE candidates, the low quality (Q=1) LAE candidates, and 
the known low-$z$ interlopers. The coadded spectrum of each set 
of galaxies was fit with a Gaussian, with the goodness of fit parameterized by the 
reduced $\chi^{2}$. As expected, the Gaussian model does a poor job at reproducing the 
observed line profile for the high quality LAE candidates ($\chi^{2}/\nu = 2.38$). Conversely, 
both the low quality LAE candidates ($\chi^{2}/\nu = 0.67$) and the low-$z$ interloper 
($\chi^{2}/\nu = 0.88$) population are statistically well fit by the Gaussian profile. Despite the statistical 
significance of the fit, visual inspection of the low-$z$ interloper population shows the line profile to 
be clearly more symmetric than the low quality LAE candidates, as should be the case if at least 
some of the low quality LAE candidates are real. Additionally, the best-fit Gaussian to the low quality candidates has a FWHM of 0.68 \AA, nearly twice as large as the best-fit 
FWHM of the known low-$z$ interlopers, further suggestive that the lower quality LAE population contains at least some genuine LAEs. The inverse of the asymmetry parameter (Section 3.1) was 
also calculated for the luminosity-weighted co-added spectra of each of the galaxy subsets, with 
values of 0.35, 0.57, and 1.14 for the high-quality LAE candidates, low quality LAE 
candidates, and low-$z$ interloper population, respectively. Both of these results reinforce 
the conclusions reached from analyzing the individual spectra: that the 
high quality LAE candidates probably represent a real population of LAEs while the lower 
quality candidates probably represent some combination of genuine high redshift LAE galaxies 
and low-$z$ interlopers. The results of these calculations did not change significantly 
if we instead use uniform weighting. 

\section{Properties of the Cl1604 Ly$\alpha$ Emitters}

\subsection{Photometric Limits}

The broadband photometry associated with the Cl1604 data set was designed almost exclusively to 
select spectroscopic targets for the supercluster at $\langle z \rangle = 0.9$, sampling down to 3$\sigma$
limits of 24.8, 24.3, and 23.6 in $r\arcmin$, $i\arcmin$, and $z\arcmin$ respectively. 
These magnitudes were calculated by measuring the magnitude of a circular object with a 1$\arcsec$ diameter, where each pixel has signal equal 
to three times the sky rms (effectively a circular top hat profile). An aperture of 1$\arcsec$ was chosen to match the average 
seeing conditions from on Palomar mountain during our observations. 

The depth of these observations are only sufficient to probe the continuum luminosities of the most massive 
galaxies at high redshift ($z \geq 4.4$). Indeed, only one of our LAE candidates (16XR1.72, an 
object that was subsequently picked as a spectroscopic target) was detected to the depth of these 
images. The accompanying archival Suprime-cam observations have a 3$\sigma$ limiting magnitude V$\sim$24.0 for the 
same choice of aperture as the other ground-based images. The exact value of this limit is unknown due to imperfect photometric calibration, 
though it is probably accurate to $\sim$0.2 mags based on comparisons between the measured Subaru magnitudes 
and overlapping fields with precise photometric calibration. 

The ACS observations are significantly deeper, reaching 3$\sigma$ limits of 
26.1 and 25.5 in F606W and F814W in most of the pointings and 26.8 and 26.3
in two deeper pointings centered on clusters A and B. Photometric limits in the ACS pointings
are calculated for a 0.3$\arcsec$ circular aperture using the same method as the ground based limiting magnitudes. A smaller 
aperture was chosen because of the significant increase in resolution ACS provides relative to the ground based images.  
These 3$\sigma$ limits are conservative limits on the depth of our images as the differential number counts do
not turnover (hereafter ``turnover magnitude") until magnitudes that are 0.1-0.2 fainter than the 3$\sigma$ limits of the ground-based data 
and 0.5-1 mags fainter than the limits of the ACS data. 
Even though the ACS data does not overlap the entirety of our spectral coverage, only two of our 17 LAE candidates (SC2NM1.45 and SC2NM2.61) fell 
outside the ACS area. Despite this, only three of the 15 LAE candidates that were covered by ACS pointings were detected in the ACS 
imaging. 

In order to place limits on the broadband photometry in the absence of detections, 
local versions of $3\sigma$ limiting magnitudes were measured for each LAE candidate from the data using a method similar to the 
measurement of the 3$\sigma$ limiting magnitudes for each image. However, rather than measuring the rms over 
a large portion of the image, the rms was instead measured in a statistically significant region either at 
the central location of the galaxy (inferred from the spectroscopy, assuming the object was at the center of 
the slit) or near the target location if the object was superimposed spatially with the target. For $3\sigma$ limiting 
magnitudes in the ACS images, this rms value per pixel (corrected for correlated noise from pixel 
subsampling) was multiplied by the number of pixels covered by an object with a circular aperture 
of radius 0.21$\arcsec$. This number was motivated by the half light radius of LBGs 
(Steidel et al. 1996b; Ferguson et al. 2004) and intentionally designed to 
overestimate the limiting magnitude of such objects; all of the LAE candidates detected in the 
ACS imaging had detected magnitudes significantly dimmer than the corresponding $3\sigma$ limiting magnitude. In addition, the 
$3\sigma$ limiting magnitudes were measured with the Palomar LFC imaging 
using similar techniques. As before, a circular aperture of $1\arcsec$ was used in the LFC calculation, as a typical LAE would not be appreciably 
different spatial extent than a point source in the LFC images. Table \ref{tab:nondettable} gives the $3\sigma$ limiting 
magnitudes of all the non-detected LAE candidates as derived from both the ACS imaging (when available) and the Palomar LFC imaging.

\subsection{Equivalent Widths}

The EW is typically 
calculated for the Ly$\alpha$ line in the following way (Dawson et al. 2004):

\begin{equation} 
EW(Ly\alpha) = \frac{F_{Ly\alpha}}{f_{\lambda}(1+z),}
\label{eqn:EW} 
\end{equation}

\noindent where $F_{Ly\alpha}$ is the total line flux in the Ly$\alpha$ line and $f_{\lambda}$ is the flux 
density redward of the Ly$\alpha$ emission, a formalization that is convenient for 
measurements of LAEs in narrowband imaging surveys. 

Without proper detections of the continuum luminosity of a majority of our
LAE candidates, calculating the EW of the Ly$\alpha$ line, something that is 
strongly dependent on the continuum luminosity, is not possible. Instead we 
calculate a lower bound on this quantity. Formally, our $3\sigma$ limiting magnitude represents 
a strict upper (brighter) bound on the continuum flux density. The uncertainty in the flux loss in the 
Ly$\alpha$ line due to the slit works in the same direction; the total line flux, minimally corrected for slit 
losses (see Section 2.4), represents a strict lower (dimmer) bound on the line flux. Thus, any calculation based on 
these numbers will represent a very conservative lower bound to the EW of the Ly$\alpha$ line 
in these galaxies. 

In order for the EW measurement, or a lower bound to this measurement, to  
characterize the intrinsic properties of high-redshift LAEs, it is necessary to make some 
correction for attenuation from the intergalactic medium (IGM). This attenuation occurs primarily 
due to resonant scattering of redshifted Ly$\alpha$ photons in intervening clouds of neutral hydrogen. As 
such, only Ly$\alpha$ photons emitted by galactic components blueshifted with respect to the bulk 
velocity of the galaxy will be affected by this dampening.  Although there can be, in principle, some 
contribution to the attenuation from intervening Helium and metal systems, 
such contributions are typically small in comparison (Madau 1995). The attenuation to the blueward flux 
solely from intervening HI regions was characterized most recently by Meiksin (2006), where the fraction of 
attenuated Ly$\alpha$ photons blueward of 1215.7 \AA\ was given as:

\begin{equation} 
f_{att,Ly\alpha,b} = 1-exp\left(-5.8\times10^{-4} \, (1+z)^{4.5}\right) \, \, (z > 4),
\label{eqn:fatt} 
\end{equation}

\noindent where the argument of the exponential is the mean Gunn-Peterson optical depth for an object at a given redshift.
Assuming the LAE is rotationally supported such that there is no skew in the velocity components of the Ly$\alpha$
emitting HI regions, the true flux of the Ly$\alpha$ line in LAEs (for z $>$ 4) is given by:

\begin{equation} 
F_{corr,Ly\alpha} = \frac{F_{Ly\alpha}} {0.5 + 0.5 (1-f_{att,Ly\alpha,b}),}
\label{eqn:ftrue} 
\end{equation}

\noindent an expression that ignores any dust extinction of the Lyman continuum. While Equation \ref{eqn:fatt} is derived from an average of
different lines of sight from observed data, we use it here to correct on a galaxy by galaxy basis. Though making this correction may introduce 
significant bias to the EW measurement of a single galaxy, correcting our entire sample produces a distribution which more accurately reflects the 
true contribution of star-forming processes in these galaxies.  
After correcting each galaxy's line flux 
using Equation \ref{eqn:ftrue}, the upper bound of the continuum flux density, $f_{\lambda}$,
was estimated. For the bulk of our sample which went undetected in the photometry, the flux density was estimated with both 
the $3\sigma$ limiting magnitude in the band encompassing the Ly$\alpha$ emission and 
in a band just redward of the Ly$\alpha$ emission. For the higher redshift 
galaxies ($z>5.5$), we had no bands completely redward of the Ly$\alpha$ line with sufficient 
depth to make a meaningful estimate of the EW with the LFC data (see Figure \ref{fig:LYAbands}), as our $z\arcmin$ imaging 
was shallower than our other bands. Both cases the LAEs fall within the ACS imaging and we therefore use only the 3$\sigma$ magnitude limit 
in the F814W filter to estimate the flux density redward of the Ly$\alpha$ line. 

Since these galaxies are undetected in the ACS or LFC data, both the EW measurement from a band
encompassing the Ly$\alpha$ line and from a band longward of the Ly$\alpha$ line represent a reasonable approximation to the lower limit on the true EW. 
Although we formally calculate EWs from the 3$\sigma$ limits in bands containing the line, the true lower bounds to the EWs are characterized solely by the EWs calculated from 
the 3$\sigma$ limiting magnitudes of bands redward of the Ly$\alpha$ line. In addition, we have  
calculated the EWs using the turnover magnitude (see Section 4.1 for definition) in the ACS imaging. For the shallower ACS pointings this magnitude
corresponds to 26.99 and 26.68 in F606W and F814W respectively, increasing to fainter magnitudes (27.76 and 28.01 in F606W and F814W) 
for the deeper ACS pointings. These turnover magnitudes are not to be confused with the completeness limits of the images, which must be
constrained through simulations; however, the turnover magnitudes are likely a good approximation to the limit at which we are 
complete for LAE-size objects. While the EWs calculated using the turnover magnitudes are not strictly lower limits, they serve 
as more reasonable (less conservative) estimates of the lower bound of the Ly$\alpha$ EW
(see Table \ref{tab:nondettable}).

For galaxies detected in the imaging (either ACS or LFC), the calculation was much 
more direct. While any measurement of the EW still represents a lower limit (due to the 
unknown amount of flux loss of the Ly$\alpha$ line from the slit), the redward continuum flux densities 
could be calculated in a straightforward way from the observed magnitudes.  These 
measurements were done, as was the case for the undetected objects, for both the band 
encompassing the Ly$\alpha$ emission and a band redward of the line emission. The lower 
limits on these EWs are shown in Table \ref{tab:dettable}, quoted at the 95\% confidence level.  

\begin{figure}
\plotone{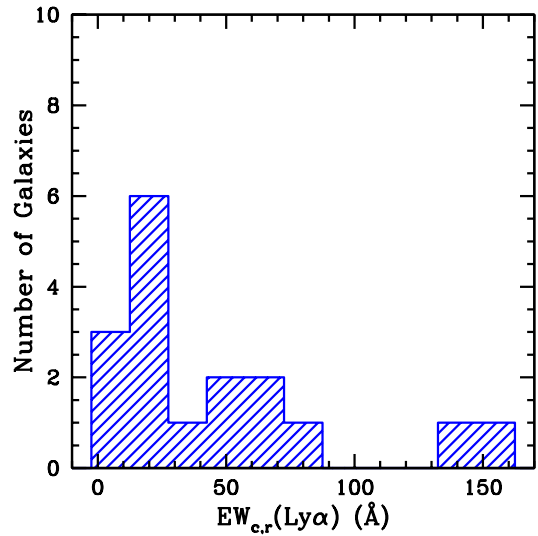}
\caption{The distribution of lower limits to the EW of the LAE candidates in our 
sample. Lower limits were calculated using line fluxes minimally corrected for slit losses and 
either the detected magnitude or the magnitude of the completeness limit in a filter completely
redward of the Ly$\alpha$ emission (for $z<4.9$) or encompassing the Ly$\alpha$ emission 
(for $z\geq4.9$).}
\label{fig:EWhist}
\end{figure}

Figure \ref{fig:EWhist} shows the distribution of our EWs calculated from either the
broadband magnitude completeness limit or the broadband magnitude
of the detection redward of the Ly$\alpha$ line.  
The distribution is strongly skewed towards very 
modest values ($\sim$25 \AA) of the EW as compared to measurements from other surveys (Dawson et al. 2004, H04, 
O08). This result is not surprising, due to the
bulk of our sample populating the faint end of the Ly$\alpha$ luminosity 
function and to the manner in which we estimate the continuum luminosity of candidates undetected in the imaging data.
Since the continuum luminosity is essentially independent of Ly$\alpha$ luminosity and since the estimated continuum luminosity 
is most likely significantly brighter than the true continuum luminosity of our candidates, the observed distribution may be more 
reflective of the way in which the EW was calculated rather than of any intrinsic properties of the LAE candidates. 
Still, the observed EW distribution is comparable 
to the results of the GLARE survey (Stanway et al. 2004, 2007), a comparably dim sample of LAEs,  
suggesting that there may be inherent properties of low-luminosity LAEs which contribute to our observed distribution.

\subsection{Star-Formation Rates and Star-Formation Rate Density}

For each LAE candidate the star formation rate (SFR) was calculated by:

\begin{equation} 
SFR \, [M_{\odot} yr^{-1}] = 9.5 \times 10^{-43} \, L_{corr}(Ly\alpha), 
\label{eqn:SFR} 
\end{equation}

\noindent where $L_{corr}(Ly\alpha)$ is the line luminosity based on the calculation in 
Section 2.4 and Equation 8.
The constant of proportionality in Equation \ref{eqn:SFR} is 
derived from the H$\alpha$ relation used in Kennicutt (1998) which assumes 
continuous star formation with a Salpeter IMF and the ratio of Ly$\alpha$ to 
H$\alpha$ photons calculated for Case B (high [$\tau$(Ly$\alpha$) $\sim10^{4}$]
optical depth) recombination (Brocklehurst 1971) for an electron temperature of 
$T_{e} = 10000$K and solar abundance. This corrected SFR still represents a lower 
limit to the intrinsic SFR of the galaxy due to the unknown amount of slit loss and due to
the fact that we do not correct for dust extinction. Tables \ref{tab:nondettable} and \ref{tab:dettable} 
list the calculated SFRs for all 17 LAE candidates. The typical LAE candidate galaxy in our sample forms stars
at a rate of 2-5 $\rm{M}_{\odot}$ yr$^{-1}$, on the low end of SFRs found in other samples.

As a consistency check, we also calculated SFRs from the rest-frame ultraviolet
continuum for the three LAE candidates that had detectable continuum in bands redward of the Ly$\alpha$ line. In no cases was this measurement 
possible from the spectra as the continuum strength in the spectra redward of the Ly$\alpha$ line was not strong enough for a reliable measurement. 
Each of the UV SFRs were derived from the flux density of the three photometrically detected LAEs, measured from their F814W magnitudes and 
converted using the Madau et al. 1998 formula:

\begin{equation} 
SFR_{UV} \, [M_{\odot} yr^{-1}] = L_{\nu,UV}/(8\times10^{27}),
\label{eqn:SFR2} 
\end{equation}

\noindent where L$_{\nu,UV}$ is a luminosity density calculated at $\sim$1500 \AA. Since the three LAEs detected in the photometric data
are at different redshifts, the effective rest-frame wavelength for the F814W filter changes slightly. However, we made no attempt to K-correct the
observed flux densities as the effective rest-frame wavelength is always within 80 \AA\ of 1500 \AA\ and because the spectrum of LAEs are relatively flat in the UV.
Table \ref{tab:dettable} lists the UV-derived SFR as well as the absolute UV magnitude for each of the three LAE candidates detected in the imaging data. The values
of the SFR derived in this manner are very similar to the SFR derived from the Ly$\alpha$ line, with the exception of one of the candidates (FG1.20), 
suggesting that slit losses for two of these candidates is minimal (not surprising since one of the objects, 16XR1.72, was targeted). 

Another quantity of interest for any 
population of high redshift galaxies is the 
star formation rate density (SFRD), which can be used to determine the onset of 
reionization in the universe. Observations of very high redshift ($z\gtrsim6$) LAEs 
and LBGs (Malhotra \& Rhoads 2004, hereafter MR04; Kashikawa et al. 2006, hereafter K06;  Shimasaku et al. 2005; Taniguchi et al. 2005;
Bouwens et al. 2003, 2004b, 2006; Bunker et al. 2004), Gunn-Peterson troughs in very high-redshift quasars (Becker et al. 2001; Djorgovski 
et al. 2001; Fan et al. 2006), and optical depth measurements from the Wilkinson Microwave 
Anisotropy Probe (WMAP; Spergel et al. 2007; Hinshaw et al. 2009) all suggest that the universe was 
reionized many 100s of Myrs prior to the observational epoch of 
our sample. The dominant population responsible for this reionization 
remains an open question. The evolution in the bulk contribution from LAEs 
has only recently been explored and shown to be a 
substantial contributor of ionizing flux from $z=3.1-5.7$ (O08; M08). However, such samples 
are only able to directly measure contributions from galaxies on the 
bright end of the luminosity function. The contribution from sub--$L_{\ast}$ 
LAEs are inferred by extrapolation of the best-fit luminosity function to faint  
Ly$\alpha$ line luminosities. 
Since many of our candidates lie far below the typical estimates for $L_{\ast}$ at these epochs,
we can better characterize the contribution from such populations. It may be that galaxies with typical Ly$\alpha$ luminosities
comparable to our sample ($\la0.1L_{\ast}-L_{\ast}$) represent the dominant contribution to the cosmic SFR amongst LAEs, 
a trend observed in $z$=2-5 LBG populations (Sawicki \& Thompson 2006).

Using the entire survey volume and all our LAE candidates, we find an SFRD of 
$5.2_{-0.6}^{+1.0}$ $\times$ $10^{-3}$ ($4.5_{-0.6}^{+0.9}$ $\times$ $10^{-3}$ excluding Q=1 candidates) 
$\rm{M}_{\odot}$ yr$^{-1}$ $\rm{Mpc}^{-3}$. These values  
should be viewed as lower limits to the SFRD, as we make no corrections for added slit loss (due to the unknown 
position of the LAE candidate) or correction for extinction of Lyman continuum photons. Though dust corrections are important to determine the  
SFRD, this correction is perhaps not problematic if we are concerned only with the number of photons available to ionize the universe 
at these epochs as any Ly$\alpha$ photon that is absorbed by dust will make little or no contribution to the re-ionization of the universe.  
The errors should also be viewed as lower limits, as we do not incorporate formal errors due to cosmic variance, though we make some effort to quantify its 
effects (see Section 5.3). Despite these complications, a comparison the observed SFRDs of the Cl1604 LAE candidates
gives values consistent with or exceeding the contributions of super-$L_{\ast}$ LAE galaxies found at $z$ = 5.7 (M08, 
O08), but significantly less than the contributions from $z=3-6$ 
LBGs ($\sim1-2\times10^{-2}$ $\rm{M}_{\odot}$ yr$^{-1}$ $\rm{Mpc}^{-3}$: 
Giavalisco et al. 2004; Bouwens et al. 2004a, 2006; Sawicki \& Thompson 2006, Iwata et al. 2007). While some LBG surveys 
correct for contributions from galaxies dimmer than the completeness limit of the survey by integrating the observed luminosity function, we 
have made no such correction. Even though this correction could, in principle, be large enough to push our observed SFRD to levels competitive with LBG surveys, 
the faint-end of the LAE luminosity function is less constrained than the low luminosity end of the LBG luminosity function making extrapolation uncertain. 
If we assume that LAEs behave like LBGs at low luminosities, exhibiting relatively shallow faint-end slopes, the contributions from galaxies dimmer than the completeness 
limit of our survey ($\sim$0.1-0.2 $L_{\ast}$) would contribute only 10\%-15\% to the total SFRD at the redshifts of interest.  

To quantify the evolution across the redshift range of our LAE candidates, we split our data 
into two redshift bins dictated by two OH transmission windows: (1)
from $z=4.1$ (the onset of our spectral sensitivity) to $z=4.95$ (the onset of 
significant contamination from bright airglow lines, see Figure \ref{fig:volume}), and (2) 
between $z=5.6$ and $z=5.8$, an atmospheric transmission 
window used by many narrowband imaging surveys. These choices of bins exclude only one 
LAE candidate, 16XR1.97 at a redshift of $z=5.02$.

The first redshift bin contains 14 LAE candidates, 11 of which are high quality. 
The volume of the survey in this wavelength range, calculated in the same manner 
as in Section 2.2, is 4.99$\times10^{3}$ $\rm{Mpc}^{3}$. The SFRD for the lower 
redshift sample is $\rm{SFRD}_{z\sim4.55}$ = $12.1_{-1.7}^{+2.6}\times10^{-3}$ 
($11.1_{-1.7}^{+2.6}\times10^{-3}$ excluding Q=1 candidates)  $\rm{M}_{\odot}$ yr$^{-1}$ $\rm{Mpc}^{-3}$, 
a density rivaling the contribution of LBGs 
at this epoch. While this bin contains a large fraction of our LAE sample, any conclusions must 
be tentative, as cosmic variance can dramatically change the observed value (see Section 5.3). 

The higher redshift bin contains two candidates (both high quality)  
within a survey volume of 1.44$\times10^{3}$ $\rm{Mpc}^{3}$. The SFRD for the higher 
redshift sample is $\rm{SFRD}_{z\sim5.7}$ = $4.4_{-1.1}^{+1.6} \times 10^{-3}$ 
$\rm{M}_{\odot}$ yr$^{-1}$ Mpc$^{-3}$, more consistent with the average SFRD 
of the survey and consistent within the errors of extrapolated SFRDs found by other 
surveys at similar redshifts (Rhoads et al. 2003, M08, O08).
Since the high redshift bin contains only two LAE candidates, the measured value of the SFRD in this bin
is highly susceptible to cosmic variance effects. 
While any conclusions that involve the high redshift bin are very uncertain, the drop in SFRD density at high redshift is 
a statistically significant effect and could possibly represent real evolution in the 
properties of LAEs as the observational epoch nears the epoch of re-ionization. If we instead choose to exclude the high redshift candidates from our sample and only use
the lower redshift bin at $z\sim4.55$, the observed SFRD is still significantly larger than those of higher redshift samples of LAEs (Rhoads et al. 2003; Ajiki et al. 2003; 
Shimasaku et al. 2006, hereafter S06; O08; M08). However, the derived SFRD at $z\sim4.55$ is also quite a bit higher than some samples at comparable redshifts (Dawson et al. 2007, S08),
suggesting that the observed change in SFRD from $z\sim4.55$ to $z\sim5.7$ probably arises through some combination of cosmic variance effects and real evolution in the LAE population.  
While marginally inconsistent with other measurements of the evolution of the LAE SFRD from $z\sim4.55$ to $z\sim6$ (O3; O8; Rhoads et al. 2003; Dawson et al. 2007),
these results are consistent with the general evolution of the star-forming properties of LBG populations 
(Bouwens et al. 2004a; Sawicki \& Thompson 2006; Yoshida et al. 2006; Iwata et al. 2007) and the overall picture of decreasing contribution to the cosmic 
SFRD from LAEs with increasing lookback time (Taniguchi et al. 2005).

\subsection{Velocity Profiles}

The observed line profile of the unsmoothed composite spectrum of the 13 high quality Ly$\alpha$ emitters was fit using a five 
parameter single Gaussian model similar to the model used in S08. In this model we assumed that the ISM absorbed all 
Ly$\alpha$ photons blueward of the centroid of an unattenuated Gaussian emission line, allowing us to produce the characteristic 
shape of the Ly$\alpha$ line. The mean wavelength of the original Ly$\alpha$ emission was allowed to freely vary, as our uncertainty 
in the redshift is coupled to our inability to quantify the extent of the absorption blueward of the Ly$\alpha$ line. 
Additionally the effective FWHM of our spectral setup, in principle a known quantity, was allowed to vary 
due to our ignorance of the placement of the LAE on the slit and the magnitude of the change 
in FWHM resolution as the LAE moves out of the slit. The background, dispersion, and amplitude of the 
Gaussian were also allowed to vary. This is, of course, a very simple model of the Ly$\alpha$ emission. In real galaxies 
there are typically multiple emission components offset in velocity space. In the case of 
LAEs, there can also be a significant excess of flux in the far red end of the line profile due to 
backscattering of Ly$\alpha$ photons by surrounding \ion{H}{2} regions 
(S08; Westra et al. 2005). Still, this model allows us to gain some insight 
into the average properties of the main velocity component of our LAE candidates.

\begin{figure}
\plotone{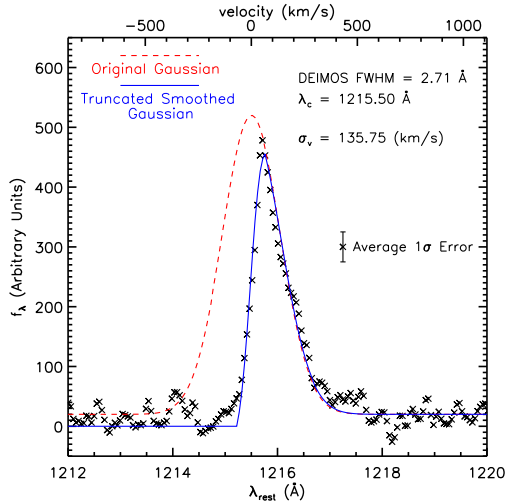}
\caption{A simple single Gaussian model fit to the unsmoothed composite (luminosity weighted) spectrum of high-quality 
(Q=2,3) LAE candidates. All flux blueward of the peak of the Gaussian has been removed
in order to approximate the effects of attenuation from HI regions. The resultant profile is smoothed with 
a second Gaussian simulating instrumental broadening. The peak wavelength and width of the original 
Gaussian as well as the instrumental broadening are all free parameters in the model. The 
model profile fits the data extremely well, with a few notable exceptions.} 
\label{fig:lineprofile}
\end{figure}

Figure \ref{fig:lineprofile} shows the best-fit model line profile overplotted on the co-added 
spectrum of the high quality LAEs. This simple model does reasonably well reproducing 
the observed line profile. It is interesting to note that if the model represents, even 
roughly, the intrinsic, unattenuated Ly$\alpha$ line, the truncation of the Ly$\alpha$ line by the 
IGM results in an attenuated line which is offset from the original line profile by nearly 100 
km s$^{-1}$. The best-fit intrinsic velocity dispersion of 136 km s$^{-1}$ is marginally consistent with the findings of S08 and LAEs 
detected in some narrowband imaging surveys (H04) and is at the extreme low end of the mass 
function of other surveys (M08; Dawson et al. 2004).

There are two main discrepancies between the data and the simple truncated 
Gaussian model. The first is the failing of the model to reproduce flux just blueward 
of the centroid of the Ly$\alpha$ line at $\sim$ 1215 \AA\ and again at $\sim$ 1214 \AA. Such excesses could 
arise from a non-trivial amount of Ly$\alpha$ photons escaping attenuation from 
pockets of neutral hydrogen. Indeed, even at the highest redshifts of our Q=3 
LAE candidates ($z\sim5.6$), Equation \ref{eqn:fatt} predicts an escape fraction 
($f_{esc}$) of $\sim5\%$, increasing to more than 40\% at the lowest 
redshifts. The model also fails to produce enough flux at the extreme red end of 
the line, showing a moderately significant decrement in flux at 
$\sim$ 1217.5 \AA\ as compared to the data. This unaccounted flux could be explained by backscattering of 
Ly$\alpha$ photons from galactic outflows as a result of star formation 
processes (Dawson et al. 2002; Mas-Hesse et al. 2003; Ahn 2004; Westra et al. 
2005, 2006; Hansen \& Oh 2006; K06). The offset between 
the observed flux excess and the centroid of the Ly$\alpha$ emission is 
$\sim$440 km s$^{-1}$, consistent with this interpretation and with 
measurements from other surveys (Dawson et al. 2002: 320 km s$^{-1}$; Westra et al. 
2005: 405 km s$^{-1}$; S08: 420 km s$^{-1}$). It is interesting to note that these signatures 
appear in both the luminosity and uniform weighted stacked spectra, suggesting that 
such outflow processes are pervasive in low-mass high-redshift star-forming galaxies.
However, both excesses are near the level of the noise in 
the co-added spectrum. While it is plausible to attribute these 
excesses to such astrophysical processes, more data are necessary to make any 
definitive conclusions. We, therefore, defer more complicated modeling of the 
composite emission line profile until all ORELSE fields are included. 

\section{Ly$\alpha$ Emitter Number Counts and Luminosity Functions}

\subsection{Number Counts and Cumulative Number Density}

The depth of our data allows us to detect galaxies down to a 
limiting luminosity of $8.8\times10^{41}$ ergs s$^{-1}$ over the entire redshift 
range of our survey and down to $3.7\times10^{41}$ ergs s$^{-1}$ at $z=4.4$. This
 is almost a factor of 2 deeper than the recent spectroscopic survey of S08, 
previously the deepest survey for LAEs to date, and nearly a factor of 10 deeper 
than the completeness limits of recent narrowband imaging surveys (O08; Dawson et al. 
2007; Murayama et al. 2007). Since our survey probes deeper in the luminosity function than 
previous surveys, any results that involve raw number 
counts of LAE candidates must be corrected for differing survey flux 
limits if proper comparisons are to be made. 

\begin{figure}
\plotone{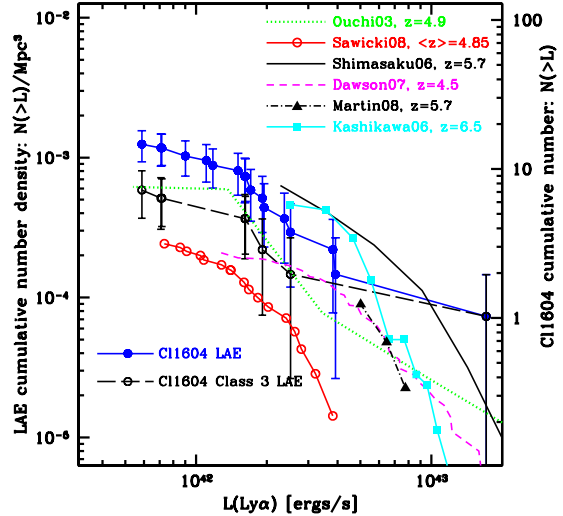}
\caption{Cumulative number density of LAEs detected in the Cl1604 field plotted as a function 
of Ly$\alpha$ line luminosity minimally corrected for slit losses ($\omega=0.8$; see Section 2.3).
The right axis shows the cumulative number of LAE candidates detected in our survey. 
The observed number counts are 
more consistent with LAE populations at higher redshift ($z\geq5.7$) than populations
at more moderate redshifts ($z\leq 4.9$). Even excluding all but our highest quality (Q=3) candidates
the number density is consistent with only the highest measurements of other surveys at similar redshifts.
Error bars are derived from a combination of Monte Carlo simulations that incorporate the uncertainties in
the luminosities (effectively which bin a given LAE candidate will fall in) and Poisson statistics, 
assuming (improperly, see Section 5.3) that LAEs have a spatial distribution reasonably similar to that of a 
Gaussian random field. Introducing formal measurements of uncertainties
due to cosmic variance (not yet possible for this sample) will cause these errors to increase. Note 
that the S08 number counts plotted in this figure and Figure \ref{fig:LAEclustercumnumdens} differ from 
those plotted in S08 as we include LAEs detected at all redshifts in their survey (with the exception of 
two marginal candidates), whereas they only include LAEs detected between $z=4.2$ and $4.9$}
\label{fig:cumnumdens}
\end{figure}

One way to disentangle number counts from the effects of varying flux limits is to 
consider the cumulative number density of LAEs. 
Since the overall shape of the function should be identical, in the absence 
of cosmic variance and any instrumental effects, the galaxy 
populations from surveys of differing flux limits can be cast in a single 
functional form. Figure \ref{fig:cumnumdens} shows both the cumulative 
number counts of all 
LAE candidates and of the highest quality (Q=3) LAE candidates in the Cl1604 field as  
compared to other surveys. Although the bulk of the 
candidates in the Cl1604 field reside at $z$ $<$ 5, the number counts 
lie above the measurements of surveys at similar redshifts. These number counts are also
significantly higher than the lower limits of S08, a survey with a nearly identical 
instrumental setup to our own. Extrapolating down to the limiting luminosity of our 
survey, the Cl1604 LAE candidates seem to be instead broadly consistent with the number 
counts of LAE surveys at much higher redshifts.  

There are many possible explanations for this discrepancy. While more exotic
possibilities are discussed in Sections 5.4 and 5.5, there are simpler 
explanations which can be explored immediately. 
One possibility that would make our results slightly more consistent with 
other surveys at similar redshifts is the exclusion of the object 16XR1.72 (the LAE
candidate at L$_{Ly\alpha}$$\sim$1.7$\times$10$^{43}$ in Figure \ref{fig:cumnumdens}). This 
is an extremely bright ($\sim$2-3 L$_{\ast}$) object that is unlikely to have been detected in 
a survey with our limited volume. Its presence is a strong indication that cosmic variance 
of such galaxies may have biased our results high for bright LAEs and its exclusion 
would serve to eliminate this effect at the brightest end of the cumulative number density distribution. 
Another possibility is the large uncertainties in the fluxes (luminosities) of the LAE candidates due to 
slit loss and flux calibration. As discussed in Section 2.3, the absolute flux calibration of
the data was accurate to only 60\%, further 
compounded by a possible systematic offset which underestimated the true value
of the flux. Ignoring for the moment the systematic offset, the uncertainty in
the flux calibration, combined with Poisson errors, could cause any given 
LAE candidate to shift nearly a factor of 2 in brightness 
(1$\sigma$) in either direction. Although significant, this shift would not change our conclusions, as our data would still be
broadly consistent with the higher redshift surveys and broadly inconsistent 
with the moderate redshift surveys. Any systematic offset 
due to slit attenuation (see Section 2.3 for a full analysis)
would cause an underestimate in the line flux and shift the cumulative number density curve to
the right, only reinforcing our conclusions. If flux errors are the only 
potential contaminant in our results, our data suggest minimal evolution in the Ly$\alpha$ number
density from $z=4.4$ to $z=6.5$. However, as we explore in the following two sections, 
there are other explanations which allow for evolution in the Ly$\alpha$ luminosity
function. 

\subsection{Clustering of Ly$\alpha$ Emitters at $z\sim4.4$ and $z\sim4.83$}

Since our survey encompasses an extremely small volume compared to
other contemporary surveys, we are susceptible to sampling galaxies whose distributions are 
unrepresentative of the distributions observed in larger LAE surveys. Effects such as
strong clustering of LAEs (O03) which lead to high levels of cosmic variance can play a large role in surveys with limited breadth. Similarly, 
variance in the observed number densities of LAE populations could be caused by inhomogeneities of intervening HI regions, 
as areas with a sparser density could manifest in increased detections of LAEs. While we will
defer any complex treatment of cosmic variance until we can include all of the
ORELSE fields, we attempt to quantify its effects in our survey in the following section. 
However, preliminary results from LAE searches in other ORELSE fields indicate similar detection frequencies, suggesting
that cosmic variance may not be the sole cause of the large LAE number densities observed in the Cl1604 field relative 
to other $z\sim$5 surveys. 

\begin{figure}
\plotone{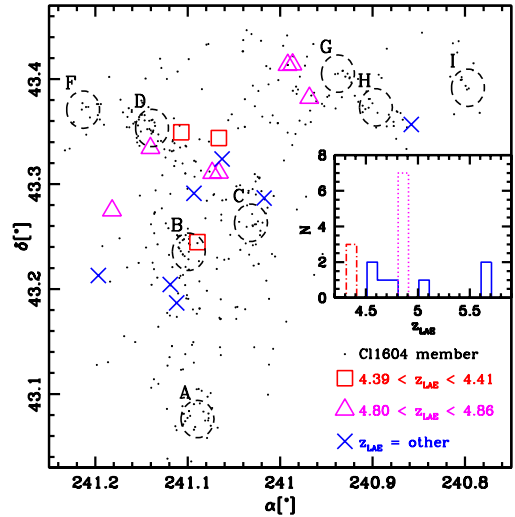}
\caption{The LAE candidates plotted against the foreground of the Cl1604 supercluster, 
with the two possible LAE structures at $z\sim4.4$ and $z\sim4.8$ shown as separate symbols.
A redshift histogram of all LAE candidates (more finely gridded than the one in Figure 
\ref{fig:zhist}) is shown to the right of the supercluster. For illustration, the name of 
each cluster as well as dashed circles that represent radii of 0.5 $\rm{h}^{-1}$ Mpc are overplotted. 
The dot-dashed and dotted lines in the histogram show the redshift distributions of the 
possible LAE structures at $z\sim4.4$ and $z\sim4.8$, respectively. The solid line shows the 
redshift distribution of the remaining ``field" LAE candidates.}
\label{fig:3dLAEplot}
\end{figure}

We do see significant evidence for clustering in our data, so it is 
possible that this clustering, combined with strong (Poissonian) sample variance, could be enough to 
explain the observed discrepancy in LAE number counts in the Cl1604 field relative to other 
surveys at the same redshift. As shown in Figure \ref{fig:zhist}, there is a
clear redshift peak at $z\sim4.8$, as well as a less pronounced
peak at $z\sim4.4$. This is recast in Figure \ref{fig:3dLAEplot} where each LAE 
is plotted against the foreground of the supercluster, with the two redshift 
peaks being differentiated from the general LAE population. 

\begin{figure}
\plotone{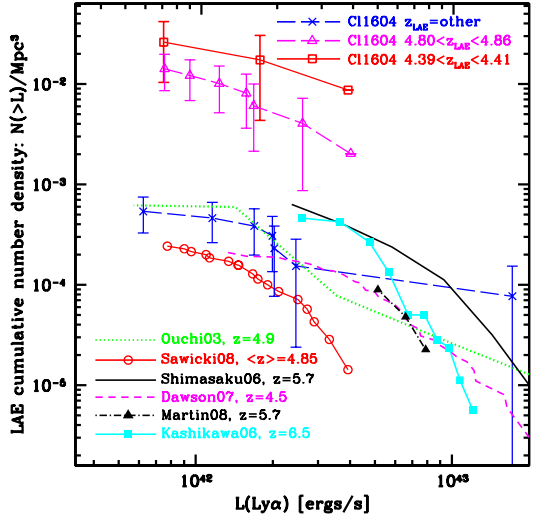}
\caption{The cumulative number density of LAEs as in Figure \ref{fig:cumnumdens}, but with the 
two possible LAE structures at $z\sim4.4$ and $z\sim4.8$ differentiated from the general population.
The volume for each potential structure is bounded by the redshift range of the LAE candidates 
in the sample and calculated from the slit area exposed at those redshifts (wavelengths). The number 
densities of these potential structures are several orders of magnitude above field populations at all
redshifts. If these structures are real, the number density of our remaining LAE candidates begin to 
be more consistent with moderate redshift ($z\leq4.9$) LAE populations. Error bars are derived in the same 
way as Figure \ref{fig:cumnumdens}. Error bars in the last bins of the two LAE structure curves have been 
removed for clarity.}
\label{fig:LAEclustercumnumdens}
\end{figure}

While the galaxies are not strongly clustered spatially, the observed distribution is broadly consistent with
the spatial distributions and number densities of other structures found at high redshift (e.g., Shimasaku et al. 2003)
and represents a significant overdensity when contrasted with
normal ``field" populations of
LAEs. Using the coverage provided by the slits to calculate the survey 
volume at these redshifts, we recover a cumulative number density of 
$2.6\times10^{-2}$ $\rm{Mpc^{-3}}$ and $1.4\times10^{-2}$ $\rm{Mpc^{-3}}$ for the 
$z\sim4.4$ and $z\sim4.83$ structures, respectively, which are inconsistent 
at the $>$99.99\% C.L. with field counts of LAEs (see Figure \ref{fig:LAEclustercumnumdens}). 

Since we have truncated the bounds of the data knowing the
redshift range of the two structures involved, it is possible that we are 
overestimating the number density for these
structures. However, in the absence of clustering we
would expect our data to closely resemble the DEIMOS response function in areas
absent of bright airglow lines. Since we are sensitive to a level far below
typical estimates of $L_{\ast}$ across the entire redshift range
of the survey, there should be little dependence of the number counts in our data on epoch. 
We are most sensitive to the redshift range 4.5-4.9 where the
DEIMOS throughput is maximized, the airglow lines are minimal, and the full survey area
is exposed to these wavelengths. It is surprising, therefore, that nearly all of 
our candidates in this redshift range lie in a narrow peak in redshift space, and that another peak
exists outside the range of this area of high spectroscopic sensitivity. 

If we exclude the LAEs in these two redshift peaks from our general LAE sample on the grounds that they belong to 
rare structures, we are left
with a total of seven (six high quality) LAE candidates in our sample that
represent a typical sampling of the LAE field populations. This reduction in the
number of candidates drops our number counts to levels consistent with, but still higher than,
other surveys at similar redshifts within the bounds of reasonable (Poissonian) sample variance (see Figure \ref{fig:LAEclustercumnumdens}).

\subsection{Cosmic Variance}

Since the LAE candidates detected in the Cl1604 field represent one of the 
first detections of a reasonably large sample of faint LAEs, little is known
about the clustering behavior of such galaxies. While the similar depth of the spectroscopic data 
in other ORELSE fields will allow for the study of statistical properties of faint LAE populations, 
the data presented in this paper is limited to one field. Thus, it is possible, given the clustering observed of 
brighter LAEs (O03; Shimasaku et al. 2003; S06), that 
our measurements of cumulative number densities (Section 5.1) or luminosity functions (Section 5.4) are 
biased by uneven sampling of the spatial distribution and the clumpiness of the population. As large statistical samples 
of faint LAEs do not exist, we must estimate by other means the likelihood that we would have detected the same number of LAE candidates if
our survey had observed a different region of the cosmos. 

In order to estimate this likelihood and to measure the magnitude of cosmic variance on the Cl1604 LAE candidates, simulations were run on 
four different samples of LAEs. The four fields were chosen because they contain a large number of LAEs, which were uniformly (or nearly uniformly) 
sampled over a large comoving volume and spanned the redshift range of the Cl1604 LAE candidates. The samples were: 1) the LALA spectroscopic
sample in the Bo\"{o}tes field at $z\sim$4.5 (D07), 2) the Subaru Deep Field (SDF) at $z\sim$4.9 (O3; hereafter SDF+LSS), 3) the SDF at $z\sim$4.9, excluding the
volume containing the large scale structure as defined in Shimasaku et al. (2003; hereafter SDF-LSS), and 4) the SDF at $z\sim$5.7 (S06; hereafter SDFhighz). 
The LAE candidates in all fields were selected using narrowband imaging methods and, in some cases, followed up with spectroscopy. Although each survey sampled a 
large volume ($\sim$1$\times$10$^{6}$ Mpc$^3$), their coverage was concentrated at nearly one epoch. However, since each survey has moderately large coverage
in the transverse directions, the variance observed in each sample likely represents a reasonable estimate for the cosmic variance of brighter
LAEs at that epoch. For simplicity, the transmission of the narrowband filter was assumed to be a top-hat response, with the hat size equal to the FWHM 
of the true filter response curve centered around the effective wavelength. 

\begin{figure}
\plotone{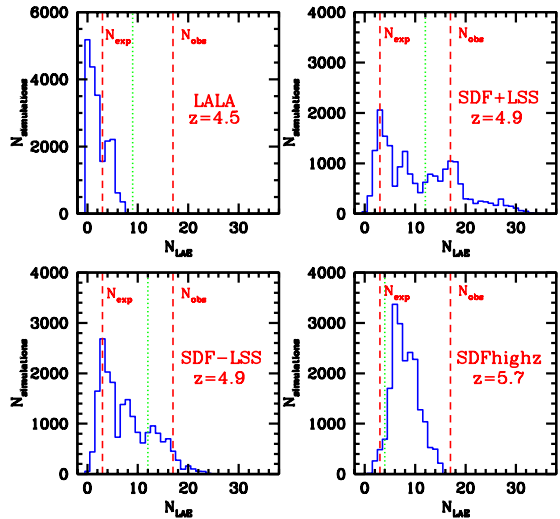}
\caption{Four different simulations of the effects of cosmic variance on the observations in the Cl1604 field. The four simulations 
are designed to cover a large range of redshifts and LAE samples. Each panel represents simulated observations of LAEs from
different narrowband imaging surveys (see Section 5.3 for details on the surveys). A histogram of the number of LAEs recovered 
in each realization is plotted in each panel. The two dashed lines correspond to the number
of LAEs expected in the Cl1604 data sampling an ``average" field at $z\sim$5 (N$_{exp}$) and the number of actual detections (all qualities) of LAE candidates 
in the Cl1604 data (N$_{obs}$). The dotted line in each panel represent the number of LAEs that we would have detected in the Cl1604 field if we instead adopt the
completeness limit of each survey. These results suggest that cosmic variance is a major contributor to the excess of LAEs described in $\S$5.1.}
\label{fig:cosmicvariance}
\end{figure}

For each realization of the simulation in each field, an area was ``observed" that would yield the survey volume of the Cl1604 spectral data (i.e. 
1.365$\times$10$^4$ Mpc$^3$) given the filter setup. For the SDF samples, where LAE candidates were selected 
using only one narrowband filter, this area corresponded to roughly 65 arcmin$^2$. In the LALA field the area observed in each simulation was 
significantly less ($\sim22$ arcmin$^2$) due to 
the continuous coverage of their five narrowband filters, which span the line of sight direction from $z=4.37$ to $z=4.57$. The observation in each 
realization consisted of counting the number of LAEs detected in a continuous square area, whose central R.A. and decl. were
determined by randomly drawing from a uniform distribution bounded by the spatial coverage of each survey. The results of these simulated observations 
are shown in Figure \ref{fig:cosmicvariance}.
 
The results of each simulation are considerably different, suggesting that, even in surveys for LAEs that probe large comoving volumes, cosmic variance
can play a large role or, alternatively, that the clustering statistics of LAEs evolve rapidly between $z\sim$4.5 and $z\sim$5.7. The variations may also 
arise from the difference in the parameters of each survey (e.g., limiting magnitude, completeness, purity). The results 
using the LALA sample in the Bo\"{o}tes field likely represents a lower limit to the cosmic variance and simulated LAE number counts because only 60\% of LAE 
candidates selected by narrowband imaging were targeted by spectroscopy. Conversely, both the SDF+LSS and SDF-LSS samples have estimated purities of 60\%-70\% (O03). 
Thus, they represent an upper limit to the cosmic variance. The SDFhighz sample, which consists of a mix of spectroscopically confirmed LAEs and objects selected 
solely through narrowband imaging, falls somewhere in between the $z\sim4.9$ SDF and the LALA sample in terms of completeness and purity.

In each panel of Figure \ref{fig:cosmicvariance}, the two dashed lines mark the number of LAEs that we expect to observe in the Cl1604 
field (i.e., 3) based on the extrapolating the number counts of surveys of LAEs at similar epochs (D07; S08) to the limiting luminosity 
of our data and the 
actual number of LAE candidates (i.e., 17) that we detect of all qualities. For both the LALA and SDFhighz fields, these simulations rule out cosmic 
variance as the sole cause of the observed excess of LAEs detected in the Cl1604 field at $>99.99$\% C.L. However, since the bulk of our galaxies lie between
$z$=4.4 and $z$=4.9, and since the LALA data are sparsely sampled, we focus on the results of the two SDF fields 
at $z\sim$4.9. In both cases, the observed number of LAEs detected in the Cl1604 are allowable within the bounds of the simulated cosmic variance. 
The likelihood of recovering at least 17 LAEs in the SDF+LSS simulations is 26\% compared to only 6\% in the SDF-LSS
simulations. This result strongly supports the conclusion reached in $\S$5.2: we may be observing at least one large scale structure of LAEs in the Cl1604 field.

However, each LAE sample from which these simulations were drawn is, on average, significantly brighter than the Cl1604 LAE candidate population. The 
completeness limits of each survey are roughly 3-8 times brighter than that of the Cl1604 survey. Thus, many of the Cl1604 LAE candidates may 
not be detected in these surveys. In order to place a lower bound on the number of galaxies that would have been detected (assuming all
of our LAE candidates were at the survey redshift), we cut the Cl1604 LAE candidate population at Ly$\alpha$ luminosities at or above the 
luminosity corresponding to the completeness limit of each survey. This number, plotted as a dotted line in each panel of Figure \ref{fig:cosmicvariance}, is a lower limit 
since the luminosities calculated for the Cl1604 LAE candidates are lower limits. Since we are cutting at the completeness limit and not the limiting luminosity, 
the number of LAEs that would be detected by each survey had they observed the Cl1604 field and covered a volume equivalent to the Cl1604 survey volume 
lies somewhere between the dotted and the rightmost dashed line in each panel. Including this cut, we 
find that the hypothesis that our field contains a large-scale structure is still favored, though less strongly, as 44\% of the simulations in the SDF+LSS 
field recover 12 or more LAEs (the number of Cl1604 LAE candidates above the O03 luminosity limit) compared to only 25\% in the SDF-LSS field. 

While it is unclear what adding in fainter LAEs to these simulations would do to the measurement of the cosmic variance, it has been observed that low-luminosity 
($\sim$0.3-0.5L$_{\ast}$) LAEs are less strongly clustered than brighter ($\gtrsim0.5$L$_{\ast}$) LAEs (O03). While this trend may not extrapolate down
to the limiting luminosities of this survey or may be an effect unique to the SDF at $z\sim$4.9, the consequences of adding in the population may be limited and 
may even serve to dilute cosmic variance. While no definitive conclusions can be reached, the main result of these simulations is 
that cosmic variance may be solely responsible for the observed excess of LAE number density detected in the Cl1604 field. Determining how likely that is, 
however, is beyond the ability of these simulations.

\subsection{Ly$\alpha$ Emitter Luminosity Function at $\langle z \rangle=4.85$}

Because a large majority of our LAE candidates remain undetected in our 
imaging, these objects are equally likely to fall anywhere within the 
bounds of our slits or perhaps, depending on their brightness, outside the 
bounds of our slits. Without any way of recovering the true position of these
objects, the slit loss correction must be approached statistically. 

The statistical correction is made using the simulation code designed 
specifically for this purpose by Martin and Sawicki (see MS04 for a more
detailed explanation). Briefly, for each realization the underlying galaxy 
population is simulated by setting the parameters $L_{\ast}$, $\Phi_{\ast}$,
and $\alpha$, fully describing a unique instance of the underlying 
distribution characterized by the Schechter (1976) function:

\begin{equation} 
\Phi(L)dL = \Phi_{\ast} \, \left(\frac{L}{L_{\ast}}\right)^{\alpha} \, exp\left(\frac{-L}{L_{\ast}}\right) \, \frac{dL}{L_{\ast}}, 
\label{eqn:generalschech} 
\end{equation}
 
Data are then simulated for a grid of Schechter parameters for 
redshift slices of $\delta z = 0.1$ running from central redshifts of $z=4.15$ to $z=6.35$ and 
multiplied by the area exposed to each redshift interval in our survey. The area exposed on the sky was also allowed to vary 
(in a known way) as a function of source flux and the seeing. 
Since brighter objects can fall further from the center of the slit (widthwise) 
and still be detected by our survey, the area of our survey at all redshift intervals increased with increasing LAE flux.
The range of simulated LAE luminosities for each redshift bin was left unbounded on the bright end and truncated on the faint end by an LAE luminosity 
that would result in a flux of $1.9\times10^{-18}$ ergs s$^{-1}$ cm$^{-2}$ (ranging from $L($Ly$\alpha)=3.2\times10^{41}$ ergs s$^{-1}$ at $z=4.15$ 
to $L($Ly$\alpha)=8.9\times10^{41}$ ergs s$^{-1}$ at $z=6.45$).
For every set of Schechter parameters, each simulated galaxy is ``observed" by calculating the slit 
attenuation based on simulated slit losses for an unresolved source galaxy at a 
regularly sampled grid of positions with respect to the slit in 0.9$\arcsec$ seeing. The total number of LAE galaxies of all fluxes (luminosities) for that set 
of Schechter parameters is then recorded. Though we only include galaxies in these simulations with fluxes greater than or equal to the 
completeness limit calculated in Section 2.5, this choice results in conservative values of the Schechter parameters. 
Specifically, we underestimate the ``true" Schechter parameters since the actual completeness limit of our survey 
is brighter than the limit calculated in Section 2.5 (due to the unknown amount of slit losses). Therefore, it is likely that 
the faintest galaxies in this simulation were not detected in our survey. Thus, this choice results in more simulated galaxies 
than if we had attempted to make a correction for slit losses, requiring us to observe more galaxies in the Cl1604 
field to recover the same set of Schechter parameters.  

From this simulation we are not able to recover a unique set of Schechter parameters due to our 
ignorance of how many genuine LAE galaxies are in our data. Furthermore, since the simulations allow each LAE candidate 
to have a range of Ly$\alpha$ luminosities, we are not definitively setting the number of galaxies detected at any given luminosity.
This constraint would be essential if we were calculating specific values of $\Phi_{\ast}$ and L$_{\ast}$. Instead, we can 
only limit the range of Schechter parameters by bounding the number of simulated galaxies for 
a given $\Phi_{\ast}L_{\ast}$ by the total number 
of genuine LAE galaxies in our data. It seems reasonable that the number 
of galaxies observed in the simulation be equal to at least 13, the number of 
high quality LAE candidates in our sample. However, this number does not represent a hard lower bound, as
any real clustering in the data would not be accounted for 
in these simulations. Instead we set the lower bound as
seven (six high quality) galaxies, the number of LAE candidates that exist outside the two possible structures and 
constitute a lower limit on our LAE field population. 

The upper bound for this simulation is more 
ambiguous. The primary consideration is 
the nature of our completeness limit, the flux at which galaxies were cut in the simulation. 
We estimate that at our flux limit we do not miss 
more than two-thirds of the actual number of LAE galaxies in our data. Taking the 17 LAE candidates of all 
qualities as the upper bound on actual detections, this sets an upper bound for our 
simulations at 51 LAEs. 

\begin{figure}
\plotone{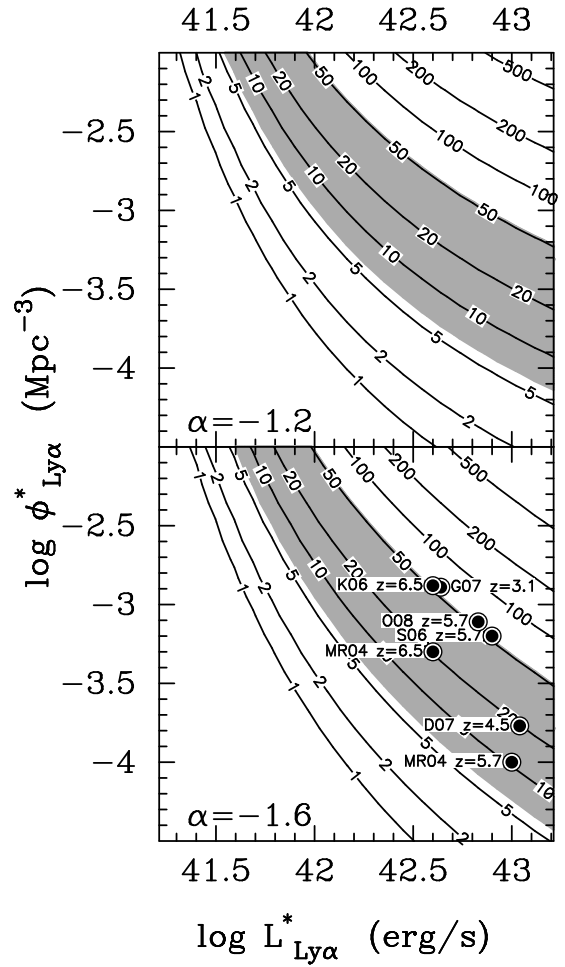}
\caption{The range of possible Schechter parameters, $\Phi_{\ast}$ and $L_{\ast}$, for simulated populations of LAEs approximating those observed in the Cl1604 field for
two different faint-end slopes ($\alpha=-1.2$, top panel; $\alpha=-1.6$, bottom panel). The data are simulated assuming a completeness limit of
$1.9\times10^{-18}$ ergs s$^{-1}$ and is corrected for statistical flux losses. Each contour through
the $\Phi_{\ast}L_{\ast}$ phase space represents the expected number of LAEs, given our instrumental setup and observing conditions, that should be
detected in our data. The shaded contour shows the phase space allowed by our LAE candidates. For comparison the values of $\Phi_{\ast}$ and $\L_{\ast}$
are shown for different surveys that used similar values of the faint-end slope. (MR04: Malhotra \& Rhoads 2004; K06: Kashikawa et al. 2006; S06: Shimasaku et al. 2006;
D07: Dawson et al. 2007; G07: Gronwall et al. 2007; O08: Ouchi 2008)}
\label{fig:Nexpected}
\end{figure}

For each panel in Figure \ref{fig:Nexpected} the shaded area represents the 
values of the Schechter parameters, the normalization ($\Phi_{\ast}$) and the 
characteristic luminosity ($L_{\ast}$) allowed by our data for the range of possible 
LAE galaxies detected by our 
survey. This analysis is done for two different faint end slopes, $\alpha$=-1.2 and $\alpha$=-1.6. 
Even if we knew the true number of LAEs in our data, due to the nature of the Schechter formalism and the sparseness of our data, 
we are not able to 
definitively determine unique values of the Schechter parameters from our simulations.  However, our data does allow us to constrain the product, $\Phi_{\ast}L_{\ast}$, 
by the range bounded by the shaded area in Figure \ref{fig:Nexpected}. 
Defining the best-fit Schechter product as the average value of $\Phi_{\ast}$$L_{\ast}$ for which we observe exactly 13 LAEs in the simulation 
and the range of possible $\Phi_{\ast}$$L_{\ast}$ values as the average of those which recover seven simulated LAEs (on the low end) and 51 simulated LAEs
(on the high end), we find a best-fit Schechter product of $\Phi_{\ast}L_{\ast}=2.2^{+3.9}_{-1.3}\times10^{39}$ ergs s$^{-1}$ Mpc$^{-3}$ for a faint-end slope of $\alpha=-1.6$.

As shown in Figure \ref{fig:Nexpected}, our observed range in $\Phi_{\ast}$ and $L_{\ast}$ is consistent with measurements made by D07, O03, and MR04 and slightly low when compared to measurements 
made by K06, S06, G07, and O08. Although there is considerable variation in the measured values of $\Phi_{\ast}$ and $L_{\ast}$ even at similar redshifts, the comparisons of the Schechter parameters seem inconsistent with the comparisons in Section 5.1, in which the cumulative number density of the Cl1604 LAE candidates were more similar to the extrapolated number counts of higher redshift surveys (K06, S06) 
than surveys at lower redshift (e.g., D07). It is possible that our conservative 
estimates of both the number and luminosity of our candidate LAEs may be the source of this discrepancy rather than any real evolution in the luminosity function of Ly$\alpha$ 
emitters from $z\sim4.85$ to $z\sim6$. Since the choice of a faint-end cutoff can severely affect simulated numbers of LAEs, especially for steeper faint-end slopes, 
our choice of a simulated flux limit of $1.9\times10^{-18}$ ergs s$^{-1}$ cm$^{-2}$ could have biased our results to lower values of $\Phi_{\ast}L_{\ast}$. To estimate the 
magnitude of this effect we ran the simulation again on a small portion of the data ($z=4.1$ to $z=4.9$) with a brighter completeness limit of $3\times10^{-18}$ ergs s$^{-1}$ cm$^{-2}$.
The recovered Schechter product was on average higher by a factor of $\sim$ 2, corresponding to a $\Phi_{\ast}L_{\ast}\approx5\times10^{39}$ 
ergs s$^{-1}$ Mpc$^{-3}$ for a faint-end slope of $\alpha=-1.6$, essentially
pushing the contours up and to the right in both panels in Figure \ref{fig:Nexpected}, encompassing the $\Phi_{\ast}L_{\ast}$ products of the other surveys within our range
of allowed values. Since the true completeness limit of the Cl1604 spectral data is somewhat uncertain (see Section 2) and since the results of these simulations are 
extremely sensitive to the choice of this limit, we are not able to distinguish between the luminosity function properties of our sample and other samples of LAEs. Instead,
we conclude that the Schechter parameters for the Cl1604 LAE population are broadly consistent with other measured values from $z\sim4$ to $z\sim6$.
 
\subsection{Weak Lensing Contributions to the Luminosity Function}

The Cl1604 supercluster is the most well-studied large-scale structure at high redshift.
While no single cluster in the structure would be considered at the high end of
 the cluster mass function (Poggianti et al. 2006, 2008; Milvang-Jensen et al. 2008; Hamana et al. 2008), with the possible exceptions of cluster A
($\sigma_{v}$ $\sim$ 703 km s$^{-1}$) and cluster B ($\sigma$ $\sim$ 800 km s$^{-1}$) (G08), the large number of moderately massive constituent clusters and the 
structure's large spatial extent make it an efficient astrophysical lens. The 
nature of the source population also lends itself to a large lensing effect: the lensing efficiency for LAEs being a 
monotonically increasing function of redshift for $z_{Ly\alpha}>z_{lens}$. The presence of this massive lens along 
with the high lensing efficiency for high redshift LAEs makes it necessary to properly account for lensing processes 
and determine whether such processes may explain the observed excess in number density counts over comparable 
field studies (see Section 5.1). 

We consider the effects of strong and weak gravitational lensing separately, with strong gravitational lensing 
effects discussed in Section 5.5. The observational weak lensing effect considered here is a 
magnification of the source population due to the lensing-induced increase in 
observed surface area while keeping surface brightness constant. Assuming the slit is sufficiently 
large (or equivalently the galaxy is sufficiently small) to encompass this increased surface area, this magnification 
increases the brightness of observed objects and the frequency of detection 
by: (1) by lensing objects into the slit which were not already within the detectable area of 
the slit and (2) by increasing the total flux of galaxies that were just below the
detection limit of the survey. Weak gravitational lensing 
could then increase both the overall normalization, $\Phi_{\ast}$, and the characteristic luminosity, L$_{\ast}$, of 
the LAE luminosity function. These effects are opposite those of all other analyses and measurement techniques
used in this study, which are intentionally designed and implemented to 
underestimate the line flux. Such effects, if significant compared to the other associated uncertainties in our measurements, 
could significantly alter our conclusions.

In principle, the most accurate approach to quantify the weak lensing effect would be to correct each LAE for the lensing-induced magnification. However,
much more spectroscopy in the field (or other similar data) would be necessary to accurately measure the effect of weak
lensing on each LAE. While a formal weak lensing analysis has been done on a small subsection of the
field around cluster A (Margoniner et al. 2005) and will be done again with newly obtained ACS
data (Lagattuta et al. 2010), the current data require us to take a more general approach.
In order to simulate the effect of weak lensing, the eight clusters that comprise the Cl1604 supercluster were modeled by
singular isothermal spheres (SIS) of the form:

\begin{equation} 
\rho(r) = \frac{\sigma_v^2}{2\pi Gr^2},
\label{eqn:sis} 
\end{equation}

While this model is an oversimplification of the true cluster mass profile, an SIS
was used in place of a Navarro-Frenk-White (NFW, Navarro et al. 1996) profile because of the closed analytic form
of the convergence and magnification solution. More importantly, however, since some of the
clusters in Cl1604 are poorly sampled by spectroscopy (clusters F, G, H, I, and J), we were unable to constrain
the characteristic radius, $r_{s}$, and the concentration parameter, $c$, needed to properly characterize an NFW profile.

Each cluster profile was simulated using velocity dispersions published in Gal et al. (2008) and 
central positions determined from the velocity and spatial centroids of the constituent cluster 
members. A velocity dispersion derived from cluster members within 1 $h^{-1}$ Mpc was adopted, chosen 
over 0.5 $h^{-1}$ Mpc or 1.5 $h^{-1}$ Mpc because it is the largest 
velocity dispersion that is relatively free from significant contamination from other clusters. Since 
$\rho$ scales as $\sigma^{2}$, this choice will allow us to measure the maximum possible (model 
dependent) lensing effect on the LAE population by the clusters. 

For each realization of the simulation, new velocity dispersions were generated for each 
cluster by a Gaussian sampling of the published 1 $h^{-1}$ Mpc velocity dispersion errors. Thus, the 
velocity dispersion of the $i$th cluster was given by:

\begin{equation}
\sigma_{v,i} = \sigma_{i,1 Mpc} + n_{i},
\label{eqn:veldis}
\end{equation}

\noindent where $n_{i}$ is sampled from a Gaussian distribution with width equal to the velocity dispersion errors. This new velocity 
dispersion, along with the static central positions of each cluster, completely 
dictated the mass map for the field of each realization; any effects from substructure, other 
structures along the line of sight, or lensing due to individual cluster galaxies were completely ignored. The source plane was created by 
averaging the $z=3.7$ and $z=5.7$ field LAE luminosity functions taken from the 
large sample from the SXDS (O08), as our survey marginalizes over any evolution in the LAE luminosity function. The resulting Schechter function is parameterized 
by $\Phi_{\ast}=5.55\times10^{-4}$ Mpc$^{-3}$, $L_{\ast} = 8.5 \times 10^{42}$, with a faint end slope of $\alpha=-1.5$.
The source population was drawn from discrete luminosity bins of width 0.1 dex, which were 
evenly spaced between L $\sim$ $10^{41}$ $L_{\odot}$ and L $\sim$ $10^{43}$ $L_{\odot}$. These limits were 
chosen to span the entire observable range of luminosities in various surveys. The number of 
galaxies in each luminosity bin was given by:

\begin{equation} 
N_{i} = \int_{L_{i}}^{L_{i+1}} \Gamma \, \Phi_{\ast} \, \left(\frac{L}{L_{\ast}}\right)^{\alpha} exp\left(\frac{-L}{L_{\ast}}\right) \frac{dL}{L_{\ast}}, 
\label{eqn:lumint} 
\end{equation}

\noindent where $\Gamma$ is the simulation volume of $3.85\times10^{6}$ Mpc$^{3}$. All galaxies in each bin were assigned the 
average luminosity in that bin, dictated by the bounds of the integration. Next, each galaxy was 
assigned a R.A. ($\alpha$) and decl. ($\delta$), generated randomly from the 
area bounded by (but not necessarily sampled by) the spectral coverage of the survey. In 
addition, a redshift was assigned to each source galaxy, drawn in a uniform random manner 
in the redshift range $z = 4.4$ to $z = 6.5$. In principle, we could have made the simulation more realistic by 
introducing evolution in the LAE population by using Schechter parameters based on various surveys at various 
redshifts. However, since the difference between the lensed and 
unlensed populations is relatively insensitive to our choice of Schechter parameters, we ignore this effect. 

Once the $\sim$ 25000 generated LAEs were assigned unique redshifts, coordinates, 
and luminosities, each luminosity was converted into an observed flux using the 
luminosity distance of each LAE. With the source plane and lens plane completely constructed, 
the lensing formalism could then be applied. For the SIS profile the shear and convergence are 
equivalent, given for the $i$th LAE as:

\begin{equation} 
\gamma_{i,j}=\kappa_{i,j} = 0.9\left(\frac{\sigma_{j,new}}{250}\right)^{2} \frac{D_{ls,i,j}} {D_{s,i} \, \theta_{i,j}},
\label{eqn:converg} 
\end{equation}

\noindent where the $j$th index represents the induced shear or convergence from the $j$th cluster in Cl1604 and $\theta$ 
is the angular separation of the $i$th LAE from the $j$th cluster in arcseconds.  
The magnification of the $i$th LAE is then calculated by:

\begin{equation} 
\mu_{i} = \frac{1}{\left(1-\kappa^{2}\right) - \gamma^{2}} = \frac{1}{1 - 2\kappa},
\label{eqn:mag} 
\end{equation}

\begin{figure}
\plotone{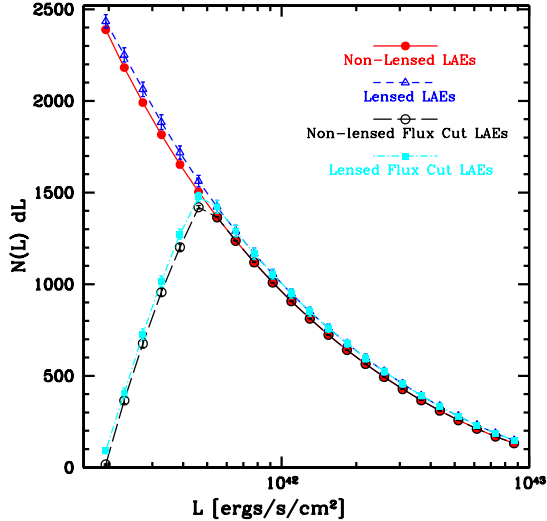}
\caption{Observations of simulated populations of LAEs in the absence of any lensing effects (solid lines) and 
after being weakly lensed by the Cl1604 supercluster (dashed lines). The two sets of curves show the differential
number as a function of Ly$\alpha$ line luminosity prior to the flux cut (continuous and short dashed lines) and subsequent to the 
flux cut (long dashed and dot-dashed lines). For all line luminosities the weak lensing number counts are consistent with the unlensed 
population within 3$\sigma$. The biggest difference between the number counts ($\sim$10\%) occurs at the 
line luminosity corresponding to the flux limit of $1\times10^{-18}$ ergs s$^{-1}$ cm$^{-2}$ and dropping to negligible 
differences (1-2\%) at higher luminosities.}
\label{fig:weaklensingnum}
\end{figure}

The flux of each LAE was then increased by its respective magnification. A flux limit of $1\times10^{18}$ ergs s$^{-1}$ cm$^{-2}$, reasonably approximating 
the flux limit of our survey, was imposed on both the original, unlensed source population and the newly 
generated weakly lensed population. The lensed flux was re-converted to an apparent luminosity, with the resulting observed 
populations shown in Figure \ref{fig:weaklensingnum}. 

The results conclusively demonstrate that the cluster-induced weak lensing effect is far too small to 
account for our increased number counts. The ensemble average increase in total detections from the unlensed to the 
lensed data is 6\%, an effect which is consistent with unlensed number counts at the 2$\sigma$ level in most of the 
bins. The effect is small regardless of the luminosity of the lensed galaxy, with an average increase in number 
counts in each bin ranging between 2\% for the brightest simulated galaxies and 8\% near the characteristic luminosity, $L_{\ast}$. 

Another way to quantify the magnitude of this effect is by the overall increase in the Schechter parameters $\phi_{\ast}$ 
and $L_{\ast}$.  While not intuitively obvious as a measurement in the overall increase (or decrease) in the 
number counts of LAEs at different luminosities, it will give us some insight into possible systematic errors (as a 
result of unquantified weak lensing effects) in our final luminosity function parameters. A Schechter parameter 
model with a fixed faint end slope ($\alpha = -1.5$) was fit to both the unlensed and lensed data using a $\chi^{2}$ 
minimization routine. As the two Schechter parameters, $\Phi_{\ast}$ and $L_{\ast}$, are degenerate with one another 
(increasing $L_{\ast}$ necessitates a decrease in the normalization in order to maintain constant numbers of galaxies), 
the quantity of interest in these fits is not the individual parameters but rather the product $\Phi_{\ast}L_{\ast}$. 
For the unlensed data the best-fit parameters resulted in a $\Phi_{\ast}L{\ast} = 4.0\times10^{39}\pm2.5\times10^{38}$, differing 
from the product of the original input Schechter parameters due to the way the simulation is coarsely binned. Errors were calculated from the
covariance matrix and the difference between the input and measured Schechter parameters. The 
fit to the lensed data resulted in a $\Phi_{\ast}L_{\ast} = 4.3\times10^{39}\pm3.1\times10^{38}$, representing an overall 
increase of 7.5\%, but also consistent within the errors to the unlensed measurements. While a correction of this magnitude 
might be important for 
precision measurements, the other uncertainties in our data (e.g., flux calibration or flux losses due 
to the slit) far outweigh any induced weak lensing signal.

\subsection{Strong Lensing Contributions to the Luminosity Function}

When a galaxy is strongly lensed, either by the cluster potential or by a massive foreground galaxy, multiple images of the background galaxy are 
created on the sky. Depending on the relative positions of the lensing potential and the source (background) galaxy as 
observed projected on the sky, the resulting images of the original galaxy can be either fainter or 
brighter than the original galaxy, thus changing the observed luminosity function.
This effect can also serve to push galaxies that would otherwise be too faint to detect above the 
flux detection limit, an effect which has been exploited by several surveys attempting to detect galaxies at very high redshift 
($z$ = 6-10, Santos et al. 2004; Egami et al. 2005; Stark et al. 2007; Richard et al. 2009).  It is also possible, though very unlikely, that 
the slit geometry is perfectly oriented to observe multiple images of the source galaxy, effectively increasing the frequency of 
LAE detections. While these effects may introduce severe bias when they occur, strong 
gravitational lensing has a comparatively small cross section, allowing us to explore the possible 
effect on the LAE candidate population in a much more direct way than the exploration of similar effects caused by 
weak gravitational lensing.  

\begin{figure}
\plotone{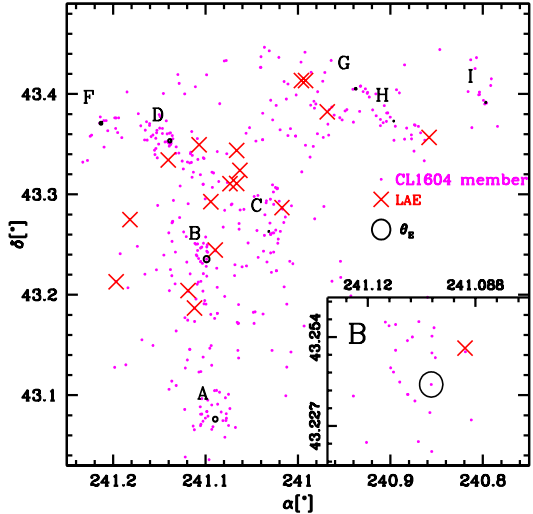}
\caption{LAEs with cluster members plotted with cluster Einstein ring radii ($\theta_{E}$), which
characterizes the onset of the strong lensing regime. Einstein radii were calculated using an 
SIS profile, with velocity dispersions
derived from all members within 1 $h^{-1}$ Mpc of the cluster center. 
All LAEs fall clearly outside the
bounds of the cluster-galaxy strong lensing regime. The galaxy closest to any cluster Einstein radius 
is shown in the bottom panel, lying many Einstein ring radii outside the center of cluster B.}
\label{fig:convergenceRAdec}
\end{figure}

For an SIS lens multiple images form only for galaxies that lie within the Einstein radius ($\theta_{E}$). 
All of the LAE candidates 
lie very safely outside of any reasonable estimate of the cluster strong lensing regime (see Figure \ref{fig:convergenceRAdec}), 
meaning any cluster-induced lensing effects would be accounted for in the previous section's 
analysis. The galaxy that comes closest to being strongly lensed is 16XR1.26, which falls greater than $3\theta_{E}$
outside of cluster B. While the true mass profile may differ from SIS and the cluster mass
may be underestimated due to our choice of velocity dispersions, it is unlikely that either of these effects would
be strong enough to increase $\theta_{E}$ to encompass 16XR1.26.  

Another issue that should be considered when discussing strong lensing of the LAE candidate 
population is any strong lensing due to individual foreground galaxies. This search 
has, by definition, a foreground galaxy companion, typically quite massive, targeted 
by the spectroscopy. Since this galaxy lies 
within the length of the small DEIMOS slit, it is possible that this effect could be 
significant. In some cases there is an additional serendipitous detection of a foreground
 galaxy on the slit which further complicates the matter. However, there are two reasons 
why we can ignore this effect. The Einstein radius for an $L_{\ast}$ galaxy 
is of the order of 1$\arcsec$. While this value will change based on the mass of the spectroscopic 
target (or other foreground serendip) and the relative redshifts of the foreground galaxy and the LAE, 
it is a reasonable estimate of where the strong lensing 
effect might be significant. All of our highest quality (Q=3) LAE candidates lie outside 
the bounds of this cutoff; any galaxy that is considered to be highly likely to be a 
LAE but is within a 1$\arcsec$ radius of either the target or a foreground serendipitously 
detected galaxy is demoted to a lower confidence class (Q=2). Even considering only Q=3 LAE candidate 
galaxies, the detection frequency in our survey is significantly higher than most other surveys 
at the redshift of our sample, an effect which cannot possibly be attributed to lensing. Furthermore, even if there 
are galaxy-galaxy lensing effects for which this analysis has failed to account,
our sample set is selected nearly identically to the LAE population detected in S08, with the 
possible exception that the spectroscopic targets in the Cl1604 field may be slightly more 
massive than those targeted by the DEEP2 survey. As they do not see similar excesses in their 
data it is likely that our observed excess comes from some combination of real, inherent properties of 
the observed LAE population, such as those discussed in the previous sections, and cosmic variance effects
and cannot be attributed solely to lensing effects.

\section{Summary}

In this paper we have described a search for LAE galaxies in the 3.214 arcmin$^{2}$ ORELSE
spectroscopic database in the Cl1604 supercluster field. In total, 17 high redshift candidate galaxies
were found in a volume of $1.365 \times 10^{4}$ Mpc$^{3}$, with 13 galaxies meeting our 
high quality criteria. The redshifts of our LAE candidates ranged from $z=4.39$ to $z=5.67$. 
Many of our candidate galaxies ($\sim$90\%) are dim compared to the typical characteristic 
luminosity at $z\sim5$, with Ly$\alpha$ line luminosities ranging from 5.9$\times$10$^{41}$ erg s$^{-1}$ ($\sim0.1L_{\ast}$) 
to 1.7$\times$10$^{43}$ erg s$^{-1}$ ($\sim2L_{\ast}$). We have contrasted our LAE candidates with a population of 
known low redshift single-emission line interlopers and blended [OII] emitters at intermediate redshifts. Our 13
high quality candidates have properties that differ significantly from the interloper population, giving us confidence
in these objects as genuine LAE galaxies. The four lower quality objects do not distinguish themselves as well from the 
interloper population, implying that these galaxies probably represent a mixture of LAEs and lower redshift objects. The increased 
frequency in LAE detections compared to other surveys demonstrates the effectiveness of LAE searches that probe deep into the luminosity 
function rather than covering large comoving volumes. Our main results are as follows:

Lower limits on the Ly$\alpha$ EW have been derived for all of our LAE candidate galaxies, 
finding a distribution peaking at EW(Ly$\alpha$)$\sim$20 \AA, similar to other low luminosity galaxies at high redshift. We have also 
derived a lower limit to the SFRs of our LAE candidate, finding that they typically form stars at a rate of 2-5 $M_{\odot}$ yr$^{-1}$. 

From the entirety of our sample we determine an 
SFRD of $4.5_{-0.6}^{+0.9}$ $\times$ $10^{-3}$ $\rm{M}_{\odot}$ $\rm{yr}^{-1}$ $\rm{Mpc}^{-3}$. This density is 
similar to or exceeding the contribution from super-L$_{\ast}$ LAE galaxies found at comparable redshifts, 
suggesting that sub-L$_{\ast}$ LAEs play an important role in keeping the universe ionized at $z\sim5$. 
Grouping our LAE candidates into low redshift ($4.1 \leq z \leq 4.95$) and high redshift 
($5.6 \leq z \leq 5.8$) bins, we find moderate evidence for negative evolution in the SFRD. We measure an 
SFRD of $11.1_{-1.7}^{+2.6}\times10^{-3}$ at $z\sim4.65$ $\rm{M}_{\odot}$ $\rm{yr}^{-1}$ $\rm{Mpc}^{-3}$ 
decreasing to $4.4_{-1.1}^{+1.6} \times 10^{-3}$ $\rm{M}_{\odot}$ $\rm{yr}^{-1}$ $\rm{Mpc}^{-3}$ at $z\sim5.7$, though
our highest redshift bin contains only two galaxies making any conclusions about the evolution of the LAE SFRD tentative. 
The derived SFRD of LAEs at $z\sim4.55$ is nearly equivalent to contributions of LBGs at similar redshifts, though this 
number is also strongly subject to cosmic variance effects.  

A simple truncated Gaussian model was fit to the composite spectrum of our high quality LAE candidates. 
The best-fit velocity dispersion was 136 km s$^{-1}$, suggesting that our galaxies lie at the low end of the 
observed LAE mass distribution. While the model fits reasonably well, there were two noticeable discrepancies 
for which the model failed to account. First, we found modest evidence for excess light at 1214 \AA\ and 1215 \AA, which 
we tentatively attributed to a non-trivial Ly$\alpha$ escape fraction. There was also an
observed excess at 1217.5 \AA\ that may be the result of galactic outflows separated from 
the LAE candidates by 440 km s$^{-1}$. As this composite represents the average  
properties of our LAE candidates, this observed excess implies that outflow processes may be 
prevalent in low mass star-forming galaxies at high redshift. 

We find the density of LAEs to be $\sim$1.5$\times10^{-3}$ Mpc$^{-3}$ for $L$(Ly$\alpha$)$\ga6\times10^{41}$ ergs s$^{-1}$, 
a frequency far higher than any other search for LAEs at comparable redshifts. We find that the excess is instead consistent with extrapolated 
cumulative number densities of higher ($z>5.7$) LAE surveys, initially suggesting minimal evolution in the LAE number density between $z=4.8$ and $z=5.7$. 

We report on the possible discovery of two structures at $z\sim4.4$ (three members) and $z\sim4.8$ (seven members). Removing these galaxies from our sample and adopting the remaining
galaxies as ``field" LAEs, we find number densities consistent with lower redshift ($z\sim4.5$) surveys, allowing for evolution in the LAE number density.

We investigate the effect of cosmic variance using simulated observations of four samples of narrowband-imaging-selected LAEs. The results of these
simulations suggest that we cannot not rule out cosmic variance as the sole cause for the observed excess in the LAE density in the Cl1604 field. 
The results of the simulations also suggest that our field contains a large scale structure of LAEs, consistent with the observed redshift 
clustering of the Cl1604 LAE candidates. 

Best-fit Schechter parameters were determined by simulating the effect of observing LAEs with our instrumental setup to
account for unknown slit attenuation. The resultant best-fit Schechter product ($\Phi_{\ast}L_{\ast}$) was found to be
$\Phi_{\ast}L_{\ast}=2.2^{+3.9}_{-1.3}\times10^{39}$ ergs s$^{-1}$ Mpc$^{-3}$. Although these simulations are sensitive to the assumed completeness limit of the survey, we
find that our results are generally consistent with other surveys both at intermediate ($z\sim4.5$) and high ($z\sim6$) redshifts.

Simulating the weak lensing effect induced by the Cl1604 supercluster using SIS models characterized by published cluster velocity dispersions, we 
find an average increase of 6\% in the observed number counts (or equivalently luminosity) of simulated LAE populations between $z=4.4$ and $z=6.5$ 
as compared to unlensed populations. The observed change in the best-fit product of the luminosity function parameters 
($\Phi_{\ast}L{\ast}$) due to weak lensing was 7.5\%, consistent within the errors to the unlensed values, and far too small to explain our observed number density 
excess. We also investigated the effects of strong lensing induced by the supercluster, finding that no galaxies are likely strongly lensed by the cluster potential. \\[14pt]

We thank Evan Kirby and Jeffrey Newman for useful discussions on coaddition techniques
and for kindly providing their codes and Nick Konidaris for helpful advice and 
guidance with DEIMOS flux calibration. We thank Raja Guhathakurta for the idea. 
D.J.L. and B.C.L. thank Matt Auger for useful discussions and criticisms. 
We also thank the anonymous referee for useful comments and suggestions. B.C.L. wishes to thank 
Eric Opland and Carl Olding for making this work possible. B.C.L. also wishes to thank Margaret Thompson
for careful reading of the paper and for the many grammar corrections:
appear up here! This material is based upon work supported by the National Aeronautics
and Space Administration under Award NNG05GC34ZG for the Long
Term Space Astrophysics Program. The research of M.S. is financially supported by the Natural Science and 
Engineering Research Council of Canada and the Canadian Space Agency. Based in part on data collected at Subaru Telescope
and obtained from the SMOKA, which is operated by the Astronomy Data Center, National Astronomical Observatory of Japan. The spectrographic data presented
herein were obtained at the W.M. Keck Observatory, which is operated
as a scientific partnership among the California Institute of
Technology, the University of California, and the National Aeronautics
and Space Administration. The Observatory was made possible by the
generous financial support of the W.M. Keck Foundation. We wish to thank the indigenous 
Hawaiian community for allowing us to be guests on their sacred mountain; we
are most fortunate to be able to conduct observations from this site.

\begin{sidewaystable*}[tp]
\clearpage
\caption{Properties of LAE candidates not detected in the imaging}
\label{tab:nondettable}
\begin{center}
\centering
\begin{tabular}{cccccccccccccccccc} \hline \hline
$ID$\footnote{ IDs are generated from a combination of mask names and slit numbers.} & $z$    & $\alpha_{2000}$\footnote{Computed assuming the LAE candidate falls in the center of the slit (widthwise).}   & $\delta_{2000}$$^{b}$ & Class\footnote{ Confidence in a candidate as a genuine LAE. 3 is most secure, 1 is least secure.}   & $F_{Ly\alpha}$\footnote{ Lower limit, calculated with a slit throughput of 0.8. Errors include a $\sim$60\% systematic uncertainty, which result from uncertainties in absolute flux measurements of DEIMOS spectra.} & $L_{Ly\alpha}$$^{d}$ & SFR $^{d,}$\footnote{ Calculated using a line flux corrected for attenuation of Ly$\alpha$ photons due to intervening HI regions.}  &$m_{F606W} $\footnote{ All ACS and LFC magnitudes are 3$\sigma$ limiting magnitudes. ACS 3$\sigma$ magnitudes are calculated using a 0.42$\arcsec$ circular aperture and LFC 3$\sigma$ magnitudes are calculated using a 1$\arcsec$ circular aperture.} & $m_{F814W}$$^{f}$ & $r\arcmin^{f}$ & $i\arcmin^{f}$ & $z\arcmin^{f}$ & $EW$    $^{e,}$ \footnote{ Rest-frame EW. Calculated using the 3$\sigma$ limiting magnitude in a band encompassing the Ly$\alpha$ line. The ACS 3$\sigma$ limiting magnitudes were used when available.} & $EW_{t}$     $^{e,}$ \footnote{ Rest-frame EW. Calculated using the turnover magnitude (see $\S$4.1) in a band encompassing the Ly$\alpha$ line.} & $EW_{r}$    $^{e,}$ \footnote{ Rest-frame EW. Calculated using the 3$\sigma$ limiting magnitude in a band redward of the Ly$\alpha$ line.} & $EW_{t, r}$    $^{e,}$ \footnote{ Rest-frame EW. Calculated using the turnover magnitude in a band redward of the Ly$\alpha$ line.} & $\lambda_{em}$\\[4pt]
& & & & & ($10^{-18}$ ergs s$^{-1}$ $\rm{cm}^{-2}$) & ($10^{42}$ erg s$^{-1}$) & $\left(\rm{M}_{\odot} \rm{yr}^{-1}\right)$ & & & & & & (\AA) & (\AA) & (\AA) & (\AA) & (\AA) \\[4pt] \hline \\[1pt]
FG2.24  &  4.40632 & 241.106979 & 43.349430 & 2 & $19.69_{-7.20}^{+11.26}$ & $3.82_{-1.40}^{+2.18}$ & $5.52_{-2.02}^{+3.16}$ & 25.24 & 24.79 & 25.63 & 25.34 & 24.03 & $>$28.7 & $>$143.4 & $>$19.0 & $>$137.0 & 6572.5\\[4pt]
GHF2.61s2  &  4.85077  & 240.996353 & 43.413406 & 3& $10.37_{-3.83}^{+5.95}$ & $2.52_{-0.93}^{+1.45}$ & $4.01_{-1.48}^{+2.30}$ & 25.28 & 24.88 & 25.72 & 25.20 & 24.01& $>$13.6 & $>$65.4 & $>$11.9 & $>$62.4 & 7112.8 \\[4pt]
SC1NM1.42 & 4.57989  & 241.111909 & 43.186732 & 2& $9.17_{-3.41}^{+5.28}$ & $1.95_{-0.72}^{+1.12}$ & $2.92_{-1.08}^{+1.69}$ & 25.36 & 24.83 & 25.00 & 25.13 & 23.39 & $>$12.1 & $>$55.1 & $>$9.5 & $>$52.6 & 6783.5\\[4pt]
SC1NM1.85 & 4.82165  & 241.073486 & 43.310581 & 1& $4.92_{-1.91}^{+2.88}$ & $1.18_{-0.45}^{+0.69}$ & $1.87_{-0.72}^{+1.09}$ & 25.40 & 24.85 & 25.69 & 25.32 & 23.89 & $>$6.7 & $>$27.7 & $>$4.8 & $>$26.4 & 7077.4\\[4pt]
SC1NM2.79 & 4.84754  & 241.066261 & 43.310909 & 1 & $3.71_{-1.43}^{+2.17}$ & $0.90_{-0.35}^{+0.53}$ & $1.44_{-0.55}^{+0.84}$ & 24.98 & 24.58 & 25.72 & 25.31 & 23.98 & $>$4.0 & $>$18.8 & $>$3.1 & $>$18.0 & 7108.9\\[4pt]
SC2NM1.34 & 5.66885  & 241.118866 & 43.204311 & 3& $4.63_{-1.80}^{+2.72}$ & $1.62_{-0.63}^{+0.95}$ & $2.93_{-1.14}^{+1.72}$ & 25.41 & 24.87 & 25.21 & 25.31 & 23.61 & $>$4.5 & $>$21.4 & ...\footnote{ No bands completely redward of the Ly$\alpha$ emission.} & ...$^{k}$ & 8107.3\\[4pt]
SC2NM1.45 & 5.62570  & 241.196610 & 43.212936 & 3& $5.56_{-2.14}^{+3.25}$ & $1.91_{-0.74}^{+1.12}$ & $3.45_{-1.33}^{+2.02}$ & ...\footnote{ Outside the ACS coverage.} & ...$^{l}$ & 26.18 & 25.22 & 23.67 & $>$6.6 & ...\footnote{Not calculated for LAEs outside the ACS coverage as the turnover magnitude was similar to the 3$\sigma$ limiting magnitude in LFC images}& ...$^{k}$ & ...$^{k}$ & 8054.9\\[4pt]
SC2NM1.78  & 4.84561 & 241.140350 & 43.334343  & 1& $6.22_{-2.38}^{+3.63}$ & $1.51_{-0.57}^{+0.88}$ & $2.40_{-0.92}^{+1.40}$ & 26.11 & 25.62 & 25.59 & 25.17 & 23.93 & $>$15.0 & $>$78.5 & $>$11.9 & $>$86.0 & 7106.5\\[4pt]
SC2NM2.61 & 4.85259 & 241.181778 & 43.274860 & 3& $2.94_{-1.19}^{+1.76}$ & $0.72_{-0.29}^{+0.43}$ & $1.14_{-0.46}^{+0.68}$ & ...$^{l}$ & ...$^{l}$ & 25.94 & 24.87 & 23.65 & $>$4.6 & ...$^{m}$ & $>$2.1 & ...$^{m}$ & 7115.0\\[4pt]
16XR1.26 &  4.39288 & 241.089600 & 43.244692 & 3& $3.68_{-1.57}^{+2.26}$ & $0.71_{-0.30}^{+0.44}$ & $1.02_{-0.43}^{+0.62}$ & 25.25 & 24.70 & 25.60 & 25.28 & 23.94 & $>$3.2 & $>$16.3 & $>$2.5 & $>$15.5 & 6556.1\\[4pt]
16XR1.97 & 5.02973 & 241.094269 & 43.292652 & 1& $8.97_{-3.38}^{+5.20}$ & $2.37_{-0.90}^{+1.38}$ & $3.92_{-1.48}^{+2.27}$ & 25.44 & 24.92 & 25.57 & 25.37 & 23.96 & $>$11.5 & $>$47.1 & $>$8.5 & $>$44.9 & 7330.3\\[4pt]
16XR2.19 & 4.53375 & 240.858124 & 43.356789 & 3& $2.84_{-1.11}^{+1.67}$ & $0.59_{-0.23}^{+0.35}$ & $0.88_{-0.35}^{+0.52}$ & 25.43 & 24.94 & 25.43 & 24.91 & 23.86 & $>$3.3 & $>$13.7 & $>$2.5 & $>$12.9 & 6727.4\\[4pt]
16XR3.26 & 4.81549 & 240.968445 & 43.382172 & 3& $6.75_{-2.57}^{+3.93}$ & $1.61_{-0.61}^{+0.94}$ & $2.56_{-0.97}^{+1.49}$ & 25.50 & 24.95 & 25.60 & 25.21 & 23.86 & $>$6.0 & $>$23.6 & $>$4.5 & $>$37.6 & 7069.9\\[4pt]
16XR3.44 & 4.69732 & 240.017355 & 43.286700 & 2& $4.90_{-2.16}^{+3.05}$ & $1.11_{-0.49}^{+0.69}$ & $1.71_{-0.75}^{+1.06}$ & 25.47 & 24.88 & 25.68 & 25.27 & 23.94 & $>$4.3 & $>$17.0 & $>$3.2 & $>$16.2 & 6926.2\\[6pt] \hline
\end{tabular}
\end{center}
\end{sidewaystable*}

\begin{sidewaystable*}[tp]
\caption{Properties of LAE candidates detected in the imaging}
\label{tab:dettable}
\begin{center}
\centering
\begin{tabular}{cccccccccccccccccc} \hline \hline
$ID$ & $z$     & $\alpha_{2000}$  & $\delta_{2000}$ & Class & $F_{Ly\alpha}$\footnote{ Lower limit, calculated with a slit throughput of 0.8. Errors include a $\sim$60\% systematic uncertainty, which result from uncertainties in absolute flux measurements of DEIMOS spectra.} & $L_{Ly\alpha}$$^{a}$ & SFR $^{a,}$ \footnote{Calculated using a line flux corrected for attenuation of Ly$\alpha$ photons due to intervening HI regions.} & SFR$_{UV}$ \footnote{Calculated from the F814W magnitude, roughly a measure of the flux density near rest-frame 1500 \AA} & M$_{UV}$$^{c}$ & $m_{F606W}$ & $m_{F814W}$ & $r\arcmin$ & $i\arcmin$ & $z\arcmin$ & $EW$    $^{b,}$ \footnote{Rest-frame EW. Calculated using the magnitude in a band encompassing the Ly$\alpha$ line.} & $EW_{r}$    $^{b,}$ \footnote{ Rest-frame EW. Calculated using the magnitude in a band redward of the Ly$\alpha$ line.} & $\lambda_{em}$ \\[4pt]  
   &  &  & & & ($10^{-18}$ ergs s$^{-1}$ $\rm{cm}^{-2}$) & ($10^{42}$ erg s$^{-1}$) & $\rm{M}_{\odot}$ yr$^{-1}$ & $\rm{M}_{\odot}$ yr$^{-1}$ & & & & & & (\AA) & (\AA) & & (\AA) \\[4pt] \hline \\[1pt]
FG1.20        &  4.40149 & 241.066400 & 43.343734 & 2 & $8.84_{-3.36}^{+5.14}$ & $1.71_{-0.65}^{+0.99}$ & $2.47_{-0.94}^{+1.43}$ & $14.0_{-1.30}^{+1.43}$ & -21.02 & 25.53 & 25.16 & 25.31 & 24.41 & ... \footnote{Not detected.} & $>$11.0 & $>$12.1 & 6566.6 \\[4pt]
GHF2.61s3  &  4.85025  & 240.991890 & 43.413297 & 2 & $16.07_{-5.88}^{+9.20}$ & $3.92_{-1.44}^{+2.24}$ & $6.24_{-2.28}^{+3.58}$ & $9.91_{-0.93}^{+0.88}$ & -20.65 & 26.89 & 25.70 & ...$^{f}$ & ...$^{f}$ & ...$^{f}$ & $>$37.3 & $>$66.8 & 7112.1\\[4pt]
16XR1.72 & 4.71230 & 241.062630 & 43.323880 & 3 & $75.42_{-27.44}^{+43.06}$ & $17.12_{-6.23}^{+9.78}$ & $26.52_{-9.65}^{+15.14}$ &  $29.75_{-0.84}^{+0.79}$ & -21.85 & 25.71 & 24.45 & 25.43 & 24.11 & 24.17 & $>$60.6 & $>$152.4 & 6944.4\\[6pt] \hline
\end{tabular}
\end{center}
\clearpage
\end{sidewaystable*}

\end{document}